\documentclass[usenatbib]{mn2e}

\usepackage{graphicx}

\usepackage{amsmath}

\def\sun{\hbox{$\odot$}}

\def\lesssim{\mathrel{\hbox{\rlap{\hbox{\lower4pt\hbox{$\sim$}}}\hbox{$<$}}}}
\def\gtrsim{\mathrel{\hbox{\rlap{\hbox{\lower4pt\hbox{$\sim$}}}\hbox{$>$}}}}

\newcommand{\mamo}[1]{\mbox{$#1$}}
\newcommand{\unit}[1]{\ifmmode \:\mbox{\rm #1}\else \mbox{#1}\fi}

\newcommand{\sbr}[1]{_{\rm #1}}
\newcommand{\expec}[1]{\mamo{\left\langle #1 \right\rangle}}
\newcommand{\mone}{\mamo{^{-1}}}

\newcommand{\kms}{\unit{km~s\mone}}

\newcommand{\mpc}{\unit{Mpc}}

\newcommand{\hmpc}{\mamo{h\mone}\mpc}

\newcommand{\ebv}{\mbox{$E(B-V)$}}

\newcommand{\ale}{\mamo{\alpha}-enhancement}
\newcommand{\secref}[1]{Section~\ref{sec:#1}}

\newcommand{\figref}[1]{Fig.~\ref{fig:#1}}
\newcommand{\tabref}[1]{Table~\ref{tab:#1}}

\newcommand{\bt}{\mamo{B/T}}
\newcommand{\br}{\mamo{B-R}}
\newcommand{\del}{\mbox{d}}
\newcommand{\rp}{\mamo{r\sbr{p}}}

\begin{document}

\title{Colours of Bulges and Discs within Galaxy Clusters and the
  Signature of Disc Fading on Infall}

\author[M. J. Hudson et al.]{%
Michael J.  Hudson,$^{1,2,3}$\thanks{Visiting Astronomer, Kitt Peak National Observatory, National Optical Astronomy Observatory, which is operated by the Association of Universities for Research in Astronomy (AURA) under cooperative agreement with the National Science Foundation.}
Jeffrey B. Stevenson,$^1$ Russell J. Smith,$^4$ Gary A. Wegner,$^5$\footnotemark[1]
\newauthor 
John R. Lucey,$^4$%
\thanks{Visiting astronomer, Cerro Tololo Inter-American Observatory, National Optical Astronomy Observatory, which are operated by the Association of Universities for Research in Astronomy, under contract with the National Science Foundation.}
Luc Simard,$^6$\\
$^1$ Department of Physics \& Astronomy, University of Waterloo, Waterloo, ON, Canada\\
$^2$ Institut d'Astrophysique de Paris - UMR 7095, CNRS/Universit\'{e} Pierre et Marie Curie, 98bis boulevard Arago, 75014 Paris, France.\\
$^3$ Perimeter Institute for Theoretical Physics, 31 Caroline St. N., Waterloo, ON, N2L 2Y5, Canada.\\
$^4$ Department of Physics, University of Durham, Durham DH1 3LE, United Kingdom\\
$^5$ Department of Physics and Astronomy, Dartmouth College, 6127 Wilder Laboratory, Hanover, NH 03755\\
$^6$ National Research Council of Canada, Herzberg Institute of Astrophysics, 5071 West Saanich Road, Victoria BC V9E 2E7, Canada}

\maketitle
\begin{abstract}
  The origins of the bulge and disc components of galaxies are of
  primary importance to understanding galaxy formation.  Here
  bulge-disc decomposition is performed simultaneously in $B$- and
  $R$-bands for 922 bright galaxies in 8 nearby $(z < 0.06$) clusters
  with deep redshift coverage using photometry from the NOAO
  Fundamental Plane Survey.  The total galaxy colours follow a
  universal colour-magnitude relation (CMR).

  The discs of $L_*$ galaxies are $0.24$ magnitudes bluer in $B-R$ than bulges.
  Bulges have a significant CMR slope
  while the CMR slope of discs is flat.
  
  Thus the slope of the CMR of the total light is driven primarily
  (60\%) by the bulge-CMR, and to a lesser extent (40\%) by the change
  in the bulge-to-total ratio as a function of magnitude.  The colours
  of the bulge and disc components do not depend on the bulge-to-total
  ratio, for galaxies with bulge-to-total ratios greater than 0.2.
  While the colours of the bulge components do not depend
  significantly on environment, the median colours of discs vary
  significantly, with discs in the cluster centre redder by 0.10
  magnitudes than those at the virial radius.  Thus while star
  formation in bulges appears to be regulated primarily by
  mass-dependent, and hence presumably internal, processes, that of
  discs is affected by the cluster environment.
\end{abstract}

\begin{keywords}
  galaxies: elliptical and lenticular, cD -- galaxies: fundamental
  parameters -- galaxies: bulges -- galaxies: clusters -- galaxies:
  photometry
\end{keywords}

\section{Introduction}

It is well-known that, at low redshifts, the centres of galaxy
clusters are dominated by red, early-type galaxies.  The
morphology-density relation dates back to \citet{HubHum31}, and was
more accurately quantified by \citet{Dre80} and \citet{PosGel84}.  The
early-type galaxies (ellipticals and S0s) follow tight scaling
relations, such as the Fundamental Plane \citep{DreLynBur87,DjoDav87}
and the colour-magnitude relation \citep[hereafter
CMR,][]{SanVis78,BowLucEll92}.  Indeed it is the existence of a tight
CMR for red galaxies that allows one to isolate a ``red-sequence''
population.

The dependence of the red-sequence CMR on environment is weak
\citep{SanVis78}, but real \citep{%
AbrSmeHut96, 
vanFraKel98, 
BalNavMor00,
CarMobBri02,
LopBarYee04, 
BalBalNic04, 
HogBlaBri04,
BarWolGra09} 
although in detail the strength of the effect depends on the
morphological selection and on the magnitude range of the sample.
These studies are based on global galaxy colours, from which it is
impossible to disentangle whether such trends might be due to, for
example, the colour of the bulge, the colour of the disc, or the
varying relative importance of bulge and disc components.

Recent spectroscopic studies of the centres of red-sequence giant
galaxies have shown that, at a given mass, those in the cores of rich
clusters are slightly older (by a few Gyr at most) than their
counterparts in lower-density environments, such as the outskirts of
clusters \citep{SmiHudLuc06}, groups \citep{ProForHau04} or the field
\citep{ThoMarBen05,BerNicShe06}.  There is little evidence of an
environmental dependence of metallicity (at fixed mass), although the
results of \cite{SmiHudLuc06} suggest that giant ($\sim L_{*}$) red
galaxies on the outskirts of clusters are $\sim 15\%$ younger and less
enhanced in $\alpha$-elements than those in the cluster cores. For
fainter dwarf galaxies ($M_{r} \gtrsim -17.5$), the environmental
dependence is stronger: \cite{SmiMarHor08, SmiLucHud09a} have shown
that the stellar ages of dwarf red-sequence galaxies in Coma vary by a
factor $\sim 3$ from the core to the virial radius.

Spiral galaxies in clusters, when compared to their counterparts in
the field, are redder \citep{Hol58, Ken83, BamNicBal09} and
``anaemic'' in appearance \citep{Van76,McIRixCal04}, and poorer in
neutral hydrogen
\citep{
DavLew73,
SulBatBot81,
HayGioChi84,
GioHay85,
CayvanBal90}.%
Several mechanisms have been proposed for the dependence of spiral
galaxy properties on environment: ram-pressure stripping of the cold
gas \citep{GunGot72}, the removal of the hot gas halo
\citep{LarTinCal80}, sometimes known as ``strangulation''
\citep{BalMor00} and tidal stripping by the cluster potential \citep{Mer84,Mam87} or by
encounters with other galaxies \citep{GalOst72,MooLakKat98}. The
former mechanisms remove the gas that is the fuel for star formation,
and so the stellar disc will dim and redden as its stars age.  This will
obviously affect the colour of the disc as well as the \bt\ ratio.  In
the tidal stripping scenario, outer parts of the stellar and gaseous
discs are removed.  A recent review of the effects of such processes
on late-type galaxies in clusters is given by \citet{BosGav06}.

From observations at higher redshifts, it is known that, in the past,
clusters contained a higher proportion of spirals compared to S0s
\citep{DreOemCou97,FasPogCou00} than present-day clusters. 
\cite{vanFraKel98} have shown that S0s are bluer and have a larger 
scatter in colour than ellipticals in a $z \sim 0.3$ cluster, 
suggesting that some of them may have recently arrived on the 
red sequence.

Clearly it is of interest to study the colours of bulge and disc
components separately, since the separate components likely have
different stellar ages and metallicities, and will be affected
differently by environmental processes.  Most previous studies have
focused on the \emph{global} colours of galaxies (sometimes subdivided
by morphological class).  Since most galaxies have both a bulge and a
disc component, this global approach makes it difficult to disentangle
the effects of the two components. There have been few previous
studies in which the separate colours of bulge and disc components have been
analyzed.  In the field, \cite{BalPel94, TerDavFro94,
  PelBal96, MacCouBel04}, among others, have studied colours of disc
and bulge components in spirals and S0s.  In clusters, two-dimensional
decompositions have been performed \citep{CaoCapRam90, Don01,
  GutTruAgu04, ChrZab04, deJSimDav04}, but the only paper to perform this
decomposition in multiple bands simultaneously is \cite{KooDatWil05},
who studied the $z \sim 0.83$ cluster MS 1054-03.

The goals of this paper are twofold. The first goal is to determine
the origin of the slope of the red-sequence CMR by decomposing cluster
galaxies into their constituent bulge and disc components, and
studying the colours of each component separately.  The second goal is
to determine how environment affects each of these components
separately, as well as how it affects the global morphology,
quantified by the bulge-to-total ratio (hereafter \bt). Other
structural parameters (bulge effective radii and surface brightnesses,
S\'{e}rsic indices and disc central surface brightness and scale
lengths) will be examined in future work.

The outline of the paper is as follows. \secref{data} describes the
sample of cluster galaxies, and details of the photometric and
redshift data. \secref{bt} describes the morphologies and
bulge-to-total ratios.  In \secref{cmr}, the total-light CMR, as well
as the CMRs for bulge and disc components is measured and its physical
interpretation is discussed in \secref{discusscmr}.  \secref{env}
quantifies the environmental dependence of the CMRs and the
astrophysical implications are discussed in \secref{discuss} .
Throughout we adopt a Hubble parameter, $h = H_0/(100 \kms \mpc^{-1})
= 1$ and a $\Omega\sbr{m} = 0.3$, $\Omega_{\Lambda} = 0.7$ cosmology.

\section{Data}
\label{sec:data}

\subsection{Cluster Sample}

As noted above, the goals of this paper are to determine the physical
origin of the colour-magnitude relation and its environmental
dependence within clusters.  In particular, we are interested not only
in the dominant red-sequence galaxies but also the contribution of
blue galaxies. Photometry for the $\sim 90$ most luminous nearby
clusters has been obtained as part of the NOAO Fundamental Plane
Survey (hereafter NFPS).  The spectroscopic component of the NFPS
\citep{SmiHudNel04} was, however, designed to follow-up only the red
galaxies.  In this paper, therefore, we focus on NFPS clusters with
complete photometry for which nearly complete deep redshift samples
are available from the literature. Eight NFPS clusters previously
studied by ENACS \citep{KatMazden98}, by \cite{ChrZab03}, by the
CAIRNS project \citep{RinGelKur03} and in the SDSS DR2 \citep{SDSSDR2}
were selected, and are listed in \tabref{clusters}.  The typical
cluster in our sample has a redshift of 14 800 \kms, a velocity
dispersion, $\sigma\sbr{cl}$ of 860 \kms, and a virial ($r_{200}$)
radius of 1.46 \hmpc, where $r_{200}=\sqrt{3} \, (\sigma\sbr{cl}/1000
\kms)\, \hmpc$ \citep{CarYeeEll97}.  The data extend to the virial
radius, $r_{200}$, and are complete within $r_{200}/2$.  Galaxies are
assumed to be cluster members if their velocities lie within $\pm 3
\sigma\sbr{cl}$ of the cluster mean.  Details of the galaxy
completeness as a function of magnitude and colour will be discussed
in \secref{comp}.

\subsection{Field Sample}
In \secref{env}, we will also compare properties of cluster galaxies
to those in the field.  In the foreground and background of the eight
clusters, there are 35 field galaxies (after exclusion of two
background clusters behind A0085 and A0151).  The median $R-$band
magnitude of the field sample is $-20.3$, very similar to that of the
cluster sample. These field data are analyzed in exactly the same way
as the cluster sample described above.

\begin{table*}
\caption{Cluster Sample. 
$N_z$ is the number of galaxies with $M_R<-19.5$.
Phot is the source of photometry: K=KPNO; C=CTIO.
}
\label{tab:clusters}
\begin{tabular}{@{}lrrrrrrrc}
\hline
Cluster &
\multicolumn{1}{c}{R.A.}&
\multicolumn{1}{c}{Decl.}&
$<cz>$&
$N_{z}$&
$M\sbr{lim}$&
$\sigma \sbr{cl}$&
$r_{200}$&
Phot\\
& 
\multicolumn{1}{c}{(J2000)} & 
\multicolumn{1}{c}{(J2000)} & 
\kms & & & \kms & \hmpc\\
 \hline
A0085 & 00 41 50.4 & --09 18 11 & 16392 & 144 &  --18.0 &  905 & 1.57 & K\\
A0119 & 00 56 16.1 & --01 15 18 & 12958 & 80  &  --19.5 &  747 & 1.29 & K\\
A0151 & 01 08 51.8 & --15 25 12 & 15769 & 51  &  --19.5 &  838 & 1.45 & K,C\\
A3128 & 03 30 23.5 & --52 32 24 & 17934 & 107 & --19.5 &  849 & 1.47 & C\\
A3158 & 03 42 56.6 & --53 38 02 & 17542 & 88  &  --19.5 &  954 & 1.65 & C\\
A0496 & 04 33 37.7 & --13 15 43 & 9717  & 127 &  --18.0 &  722 & 1.25 & K\\
A0576 & 07 21 24.2 & +55 47 02 & 11527 & 60  &  --19.5 &  871 & 1.51 & K \\
A3667 & 20 13 33.4 & --56 50 35 & 16585 & 93  &  --19.5 & 1000 & 1.73 & C \\
\hline
\end{tabular}
\end{table*}

\subsection{Photometric Data}

The photometric data used were obtained as part of the NFPS
\citep{SmiHudNel04}, from the Kitt Peak National Observatory (KPNO)
0.9 meter and the Cerro Tololo Inter-American Observatory (CTIO) 4.0
meter telescopes.  The fields of view were $59'$ x $59'$ and $37'$ x
$37'$ for KPNO and CTIO, respectively, and were centreed on the
nominal X-ray centroid of the cluster.  Exposure times were typically
400 ($R$) and 600 ($B$) seconds at KPNO, and 60 ($R$) and 150 ($B$)
seconds at CTIO.  The seeing point spread function (PSF) full-width at
half maximum typically was $\sim 1.5$ arcseconds for KPNO and $\sim
1.1$ arcseconds for CTIO.

Data were reduced with the IRAF {\tt mscred} package, and were
calibrated to the Johnson-Cousins $BR$ system using \cite{Lan92}
standard stars.  Objects and masks were generated with SExtractor
\citep{BerArn96}.

\subsection{Corrections and Completeness}
\label{sec:comp}

Redshifts of galaxies within our images were gathered from the NFPS
spectroscopic survey \citep{SmiHudNel04} and from the NASA
Extragalactic Database (NED).  Galaxies were assigned cluster
membership using an iterative $3\sigma$-clipping technique.  Apparent
magnitudes were converted into absolute magnitudes using cluster
redshift in the Cosmic Microwave Background (CMB) frame.

Galactic extinction corrections were applied using the \ebv\ map of
\citet[hereafter SFD]{SchFinDav98}, and converted to extinction using
$4.325\,\ebv$ for the B band and $2.634\,\ebv$ for the $R$-band (SFD).
K-corrections were applied to the bulge and disc components
separately, by interpolating as a function of morphological type from
the tables in \citet{FreGun94}.

\begin{figure}
  \includegraphics[width=\columnwidth]{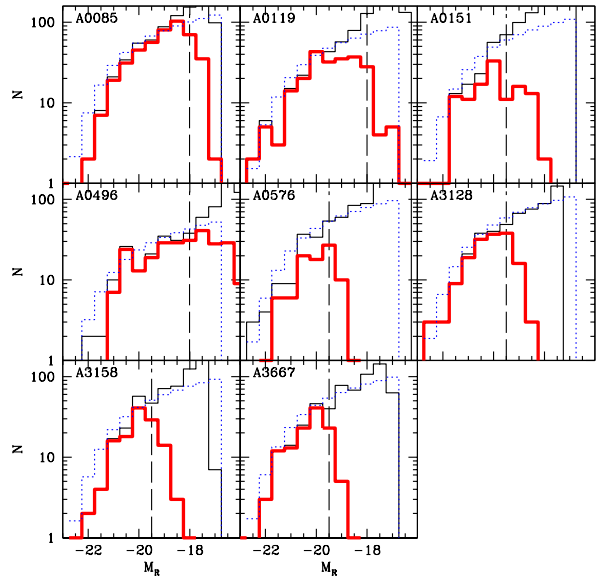}
  \caption{ $R$-band magnitude histograms for individual clusters. The
    thin black histogram is all galaxies with photometry, whereas the thick
    red histogram shows those with published redshifts. The luminosity
    functions (normalized at the bright end by assuming a Schechter
    function) are shown by the blue dotted line. In each panel, the
    vertical dashed line indicates the absolute magnitude limit
    adopted for that cluster.
\label{fig:czhist}
}
\end{figure}

\figref{czhist} shows that in general, the redshift data are complete
at the bright end ($M_{R} \lesssim -20$), but that, as expected, the
completeness decreases at fainter magnitudes.  To measure the
completeness, a Schechter luminosity function was fit to each cluster
with galaxies with $M_{R} < -21$ fixing $M^{*}_{R} = -21.14$ and
$\alpha = -1.21$ \citep{ChrZab03}, but leaving the normalization,
$\phi_{*}$, free.  The fit is shown by the dotted line in
\figref{czhist}.

Based on \figref{czhist}, we adopt a Galactic-extinction-corrected
absolute magnitude limit of $M_R < -19.5$ for our primary sample. For the
sample as a whole, the average completeness brighter this limiting
magnitude is 87\%, and the completeness in the faintest 0.5 mag bin is
66\%.  To correct for incompleteness, galaxies with redshifts are
assigned a weight that is the inverse of the probability that a
cluster member of that magnitude has a measured redshift.  Note that
the clusters A0085, A0119 and A0496 have redshift data that are
considerably deeper. These three clusters are used to define a deeper
$M_{R} < -18$ sample, which will be used later to assess correlations
at fainter magnitudes.

\subsection{GIM2D Morphological Parameters}

Bulge and disc decomposition is based on the GIM2D morphology package
\citep{SimWilVog02} which analyses postage-stamp images along with
bad-pixel masks, PSF images, and error maps and fits a PSF-convolved
model galaxy to the two-dimensional images.  The GIM2D model consists
of a infinitesimally-thin exponential disc and an axisymmetric
S\'{e}rsic bulge. The fit yields bulge and disc magnitudes, radii,
position angles, and inclinations (for discs), or ellipticities and
S\'{e}rsic index $n$ (for the bulge components).

Note that we could have chosen to fit the $B$ and $R$-band images
independently, but in that case, it would have been possible to obtain
different values in the two bands for parameters such as the radii and
position angles of given component.  In particular, because of, for
example, systematic errors in the PSF, there can be a trade-off
between the flux in the bulge and the flux in the disc components. If
such systematic errors were different in the two bands, the colours of
the bulge and disc components would not be robust.  Instead, we used
GIM2D in the two-band mode, in which simultaneous fits to the two
images are made for the radii, position angles, bulge S\'{e}rsic index
$n$, bulge ellipticity and disc inclination, while bulge and disc
fluxes are free in each filter over all pixels in the SExtractor
\citep{BerArn96} segmentation mask.  While errors of the type
described above can still affect these fits, in general, one would
expect them to be yield more robust colours for the individual
components.  

A limitation of two-band mode is that the bulge effective radii are
assumed to be the same in both $B$ and $R$ bands, and likewise for the
disc scale lengths.  In reality, disc scale lengths are likely to
differ, although for the S0s and early-type spirals studied here the
difference is quite small: \citet{Noovan07} find that the disc scale
lengths in a sample of S0-Sab galaxies are larger by $3.2\pm2.5$\% in
the $B$-band than in the $R$-band. We have simulated the impact that
this GIM2D fitting constraint may have on disc colours by fitting
model galaxies with disc scale length assumed to be different from the
true one.  We find that for a 3\% error in scale length, the
colour is biased at a level of 0.01 mags.  Thus this is a small
effect.

\subsection{Data Comparison}
\label{sec:datacomp}

\begin{figure}
\includegraphics[width=\columnwidth]{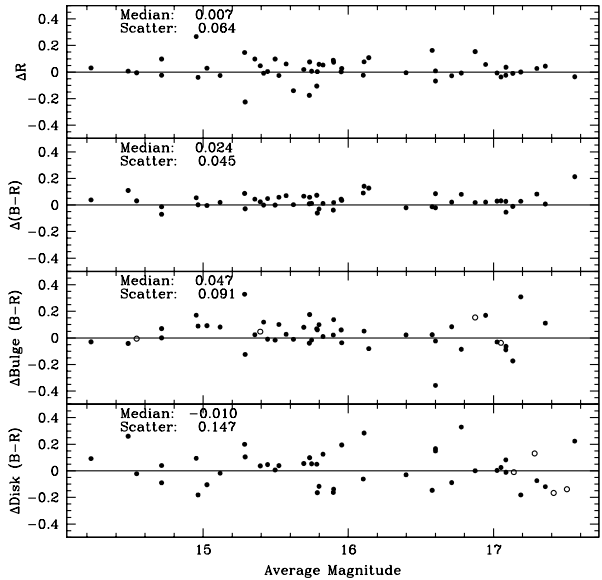}
\caption{ Comparison of the total magnitudes and colours as measured by
  GIM2D from independent runs. $\Delta$ is in the sense KPNO $-$ CTIO. 
  The upper panel compare the $R$-band magnitudes, the second panel
  shows the global \br\ colour comparison, and the bottom two panels compare
  bulge and disc colours.  The median offsets are small.  
  \label{fig:comparecolbulgedisk}
}
\end{figure}

Before proceeding with the analysis of the data, a check was performed
to ensure internal consistency of the photometric data and GIM2D
decompositions. Of the clusters in the sample, only A0151 was observed
both at KPNO and at CTIO. Figures \ref{fig:comparecolbulgedisk} shows
the comparison of the A0151 data obtained from the different
observatories, in the sense $\Delta \equiv$ KPNO - CTIO.  In the
magnitude comparisons (top panels of Figure
\ref{fig:comparecolbulgedisk}), the median offset is $0.007$
magnitudes in the $R$-band and $0.025$ magnitudes in the $B$-band.

The scatter is larger than the individual measurement errors quoted by
GIM2D. This suggests that GIM2D errors are underestimated, which is
expected since these reflect only the error due to photon noise, and
not e.g.\ seeing mismatches, sky subtraction etc.  Therefore, to the
quoted GIM2D magnitude errors, we add in quadrature a
magnitude-independent extra ``systematic'' error.
Based on the scatter in the top panel of \figref{comparecolbulgedisk},
we find that this extra error is 0.05 magnitudes.

The \br\ colour comparison is shown in the second panel of Figure
\ref{fig:comparecolbulgedisk}. The median offset between KPNO and CTIO
is $0.02$ magnitudes, which is acceptable for the purposes of this
study.  The measured scatter suggests that individual colour
measurements are accurate to 0.03 magnitudes. This is smaller than the
magnitude error in either band separately, and is due to the fact that
the fit is made simultaneously in both bands. From similar comparisons
(see bottom panels of \figref{comparecolbulgedisk}), errors in a
single measurement of the bulge and disc colours are estimated to be
0.07 and 0.11, respectively.

\begin{figure}
  \includegraphics[width=\columnwidth]{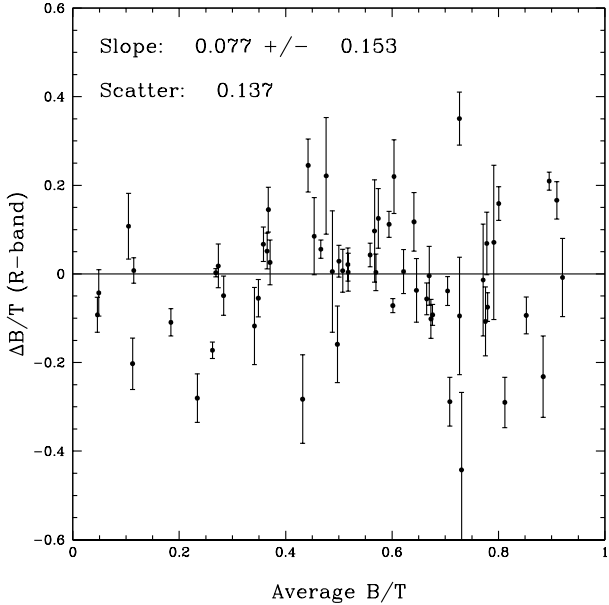}
  \caption{Difference between the $R$-band \bt\ measured from CTIO and from KPNO as a function of average $R$-band \bt.  The plotted GIM2D errors underestimate the differences, which presumably arise from other systematic sources of error such as modelling the PSF. 
    \label{fig:btrmagwerror}}
\end{figure}

Figure \ref{fig:btrmagwerror} shows the differences in the
bulge-to-total light ratios.  The observed scatter in the R-band \bt\
values is larger than the GIM2D errors. Therefore, we add in
quadrature a systematic error of $0.07$ per individual \bt\
measurement. Given the random (noise) errors, the total measurement
error in $\bt$ is then typically $\sim 0.08$.

\section{Morphologies and Bulge-to-Total Ratios}
\label{sec:bt}

\subsection{Visual Morphological Classification}

For a subset of the sample described above, postage stamp images, pixel masks and GIM2D residual images for all galaxies were inspected visually by one of us (GAW) and a visual morphological classification was assigned to each galaxy. Furthermore, cases where the automated pixel masking had failed (e.g. close galaxy pairs or missed bad columns, etc) were flagged and rejected from the sample.

\begin{figure}
\includegraphics[width=\columnwidth]{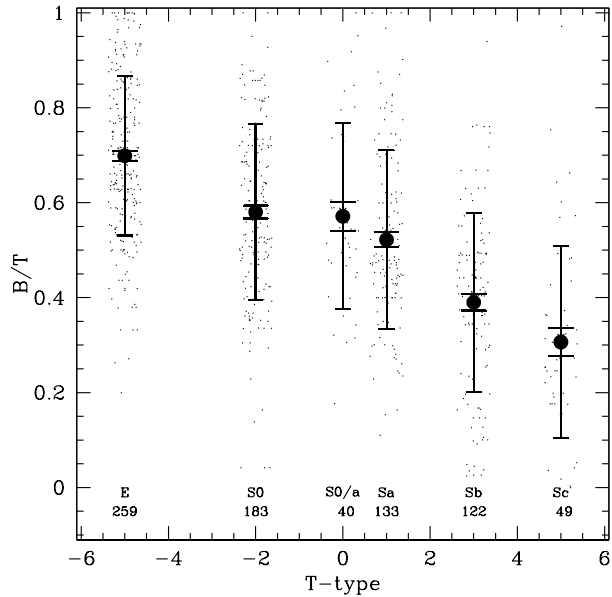}
\caption{ Visual morphological classifications encoded by $T$-type
\citep{RC3} versus R-band \bt. The filled circles show the mean \bt\ at each
$T$-type, and the long errorbars indicate the dispersion in \bt\ at
each type.  The short errorbars are the error in the mean. Small dots
are individual data points, with the $T$ slightly randomized for
graphical purposes. The number of galaxies in each bin is also
indicated. 
\label{fig:btvstype}
}
\end{figure}

\figref{btvstype} compares the R-band bulge-to-total (\bt) ratios from GIM2D with the visual classification.  The visual classifications were labeled according to the corresponding T-type and plotted against \bt. The trend of decreasing \bt\ with increasing T-type, and the dispersion in \bt\ ($\sim 0.17$) at a given type, are in generally in good agreement with previous studies \citep{Ken85, SimdeV86}, although the  median $\bt\ \sim 0.3$ for Sc galaxies is somewhat higher than expected. Since the measurement error in \bt\ is typically $0.08$ (\secref{datacomp}), the intrinsic scatter in \bt\ of a given T-type is $\sim 0.15$.

\subsection{Tests of the GIM2D decomposition}

GIM2D fits a S\'{e}rsic bulge plus disc model to the $B-$ and $R-$band
data simultaneously. While we argue above that the constraints of
these fits lead to more robust colours of the bulge and disc
components, it's not necessarily true that these components --- needed
to reproduce the 2-D light distribution --- correspond to real bulge
and disc components. For example, a pure bulge component with an
isophote twist might be fitted with a (untwisted) bulge and a ``disc''
which accommodates the twist.

\begin{figure}
  \includegraphics[width=\columnwidth]{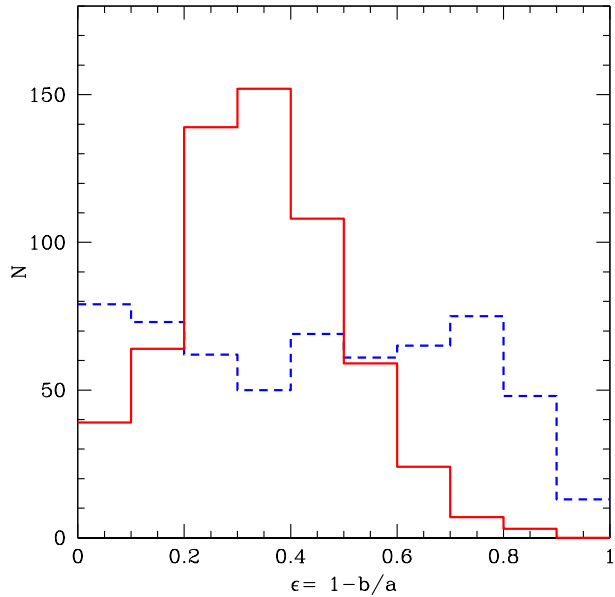}
  \caption{ Distribution of ellipticity for disc (dashed blue) and bulge
    (solid red) components.  Note that there is not an excess of disc
    components with ellipticities typical of bulges. 
    The two distributions are very close to what
    one expects \emph{a priori} on physical grounds indicating that
    the disc and bulge components are not artifacts of the fitting procedure. 
    \label{fig:ellip}}
\end{figure}

One way to check the reality of the bulge and disc components is by
plotting their ellipticity distributions. If the ``discs'' are really
related to the bulge components, they would be expected to follow the
ellipticity distribution of bulges, i.e. there should be few with
$\epsilon = 1 - b/a > 0.5$. \figref{ellip} shows the ellipticity
distribution for bulges and disc components.  If the discs are
infinitely thin, transparent and seen at random inclinations, then
they would have a uniform distribution in $\epsilon$.  The discs in
\figref{ellip} are very close to this scenario, although there is a
deficit of very high $\epsilon > 0.9$ discs. We attribute this to the
finite vertical scale-height of discs: even seen edge-on, such discs will
have $b/a > 0$.  Note that this result is different from that found by
\cite{DriPopTuf07} for discs in the \emph{field}.  In addition to the deficit of very high $\epsilon$ discs, they also find a deficit of \emph{moderately-inclined} discs: for example, in the range $0.5 \lesssim \epsilon \lesssim 0.7$ (corresponding to $60\degr \lesssim i \lesssim 73$) their counts are reduced by $\sim 40\%$ (cf. their Figure 5), whereas ours are consistent with being constant. They conclude that dust obscuration dims these moderately-inclined galaxies resulting in their under-representation in a magnitude-limited survey. In contrast, our uncorrected disc $\epsilon$ distribution looks very similar to their distribution \emph{after correction} for dust extinction.  The difference between the field and cluster disc ellipticities suggests that cluster discs may be depleted in dust, and hence considerably more transparent.  In contrast to the flat distribution of disc ellipticities, the bulge components are clustered between $0.1 < \epsilon < 0.5$, as expected based on the distribution of E+S0 ellipticities in clusters \citep[e.g.][]{JorFra94}.

\subsection{Dependence of \bt\ on Magnitude and Environment}

\begin{figure}
\includegraphics[width=\columnwidth]{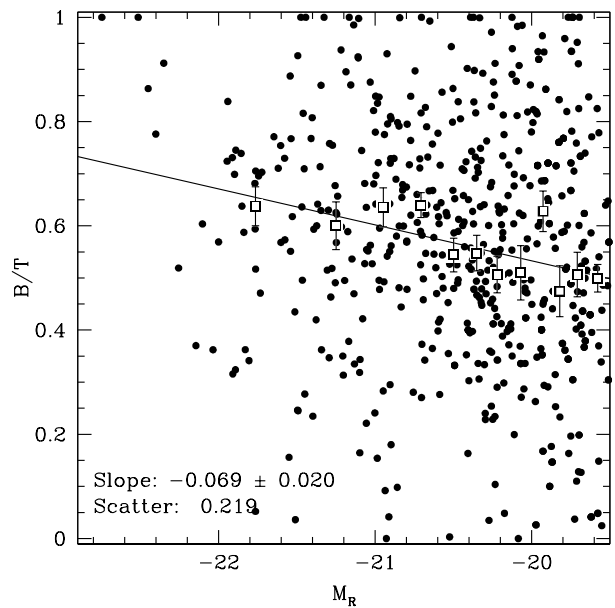}
  \caption{\bt\ in the $R$-band as a function of magnitude for the $M_R < -19.5$
    sample.  The square symbols with error bars show the median \bt\
    in a bins of magnitude. The solid line is a fit to the medians.
    \label{fig:btrmag}}
\end{figure}

Before discussing the colours of bulge and disc components, it is
interesting to study the dependence of \bt\ on magnitude and
environment. Throughout this paper, unless explicitly stated
otherwise,\bt\ refers to the $R$-band value. \figref{btrmag} shows the
weak, but statistically significant, dependence of the bulge-to-total
ratio on magnitude, with more luminous galaxies being more
bulge-dominated, mirroring the trend of morphological type with
luminosity.

\begin{figure}
  \includegraphics[width=\columnwidth]{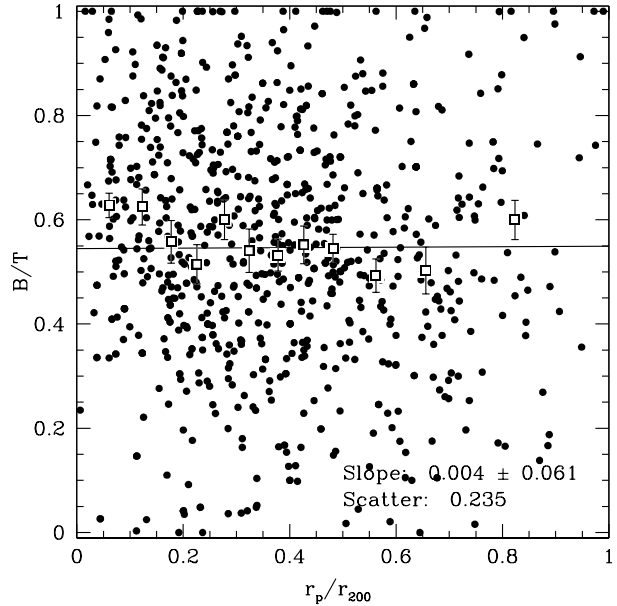}
  \caption{\bt\ in the $R$-band as a function of scaled projected radius for the
    $M_R < -19.5$ sample. The square symbols with error bars show the
    median \bt\ in a bin of $\rp/r_{200}$. The line is a fit to the
    medians.\label{fig:btrrad}}
\end{figure}

The morphology-radius relation, where morphology is quantified by \bt\ and the projected radius, $r_{p}$ is scaled by the virial radius, $r_{200}$, is shown in \figref{btrrad}.  The variation in \bt\ is rather weak: only in the very innermost regions ($\rp \lesssim 0.2 \,r_{200}$) is there evidence for a difference in the median \bt. However, a Spearmank rank correlation test does suggest a correlation at the 98.7\% confidence level. Moreover, fitting \bt\ as a function of $\log(\rp/r_{200})$ yields a significant slope: $-0.108\pm0.033$\footnote{In \figref{magrad} below, we demonstrate that there is no significant magnitude-radius relation, hence the correlation found here is not an artifact of luminosity segregation coupled with the dependence of \bt\ on magnitude.}. Thus while there is a clear difference between morphologies in clusters and in the field, the morphological segregation as defined by \bt\ \emph{within} clusters appears to be limited to the innermost regions.

\begin{figure}
  \includegraphics[width=\columnwidth]{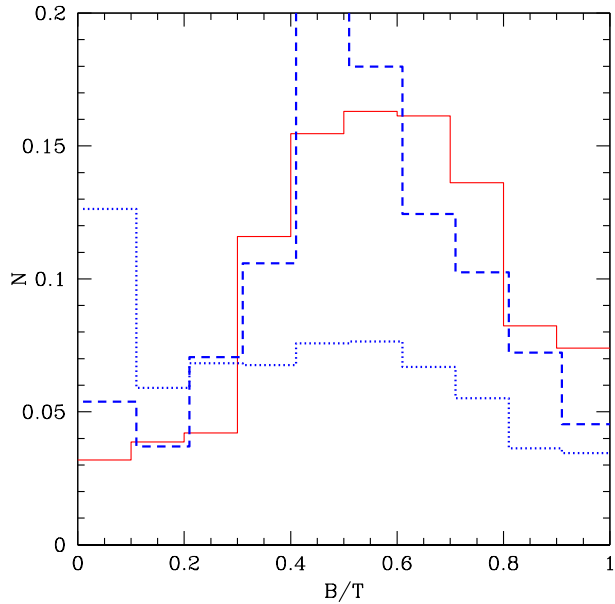}
  \caption{ 
The solid red histogram shows the distribution of $R-$band \bt\ for $M_R <
-19.5$ galaxies in the NFPS clusters, the dashed blue line is the same but
for the $B-$band \bt,  and, for comparison, the dotted blue line shows the $B-$band \bt\ for
the field sample \citep{AllDriGra06}, selected to the same magnitude limit as the cluster sample ($M_{R} < -19.5$).  The most striking differences are the excess of late-type spiral ($\bt <0.1$) systems in the field, and the excess of S0-like ($0.4 < \bt\ < 0.6$) morphologies in the cluster. 
\label{fig:bthist}}
\end{figure}

Finally, it is interesting to compare the cluster results with those in the field. The distribution of \bt\ (for $\rp < 0.5\, r_{200}$) is shown in \figref{bthist}. For comparison, we also show the $B-$ band \bt\ distribution in the field \citep{AllDriGra06} for the same absolute magnitude limit as applied to the NFPS clusters. Compared to the field, clusters contain an excess of galaxies with S0 morphology  ($0.4 < \bt < 0.6$), and a clear deficit of spirals, particularly those of the latest types, or lowest bulge fractions $(\bt < 0.1)$.  The distribution of \bt\ in the group environment \citep{McGBalHen08} is intermediate between that in the field and in clusters. 

\section{The Red-Sequence Colour-Magnitude Relation Revisited}
\label{sec:cmr}

In this section we examine the global-colour vs.\ magnitude relation
for individual clusters as well as the universal relation for all
clusters. We then decompose the global-CMR into the CMRs for bulges
and discs separately.

\subsection{Robust Fits to Galaxies within $r_{200}/2$ }

We will show below (\secref{env}) that the colours of some components
depend on cluster-centric radius.  Consequently, in this section, we
restrict fits to galaxies with projected cluster-centric radii $\rp <
r_{200}/2$.  Furthermore for fits to the red-sequence, we do not wish
to be biased by very blue galaxies, nor by outliers with anomalous
colours. Therefore, in order to fit the CMR in a robust way, the data
were first binned by magnitude, with an equal number of data points in
each bin.  For each bin the median was calculated, and a linear least
squares trend was fit to these medians.

To estimate the scatter of the data, the semi-interquartile range
(SIQR) of the data is calculated, which is then converted into a the
equivalent dispersion ($\sigma = 1.5\mathrm{SIQR}$) of a Gaussian
with the same SIQR.  In order to compare the colour-magnitude relations
with different slopes, a normalized colour is defined, $(B-R)_{*}$ the
value of the fit at $M^{*}_{R} = -21.14$. The errors on the normalized
colour were calculated using 1000 bootstraps.

\subsection{Global Colour-Magnitude Relation}

\begin{figure}
  \includegraphics[width=\columnwidth]{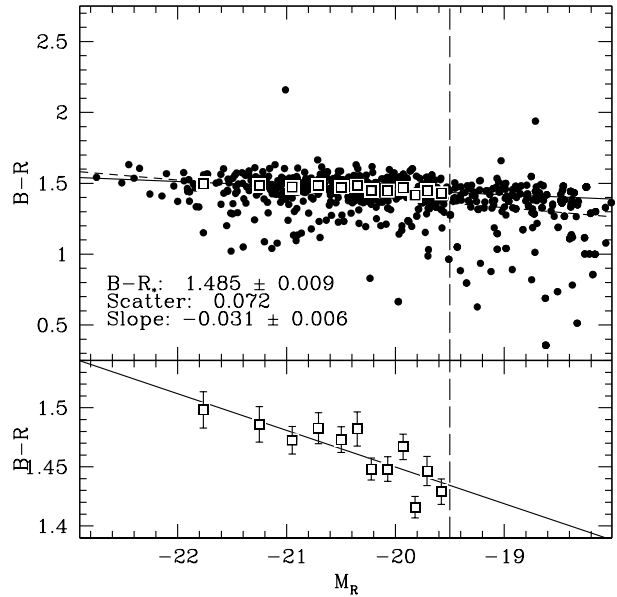}
  \caption{ The colour-magnitude diagram for all 8 clusters combined. 
    \emph{Upper panel:} 
    Filled circles indicate total-light colours of individual cluster galaxies with $\rp < 0.5 r_{200}$.
    Median colours in bins of magnitude are shown by the open squares.
   The robust fit to the median colours is shown by the solid line is to galaxies with $M_R < -19.5$ and $\rp < r_{200}/2$. 
   \emph{Lower panel:} Same median data  as in the upper panel, but shown at a different scale to better see the slope.
    \label{fig:colmag} }
\end{figure}

The global colour-magnitude relation is shown in Figure
\ref{fig:colmag}.  For galaxies with $M_R < -19.5$ the normalized
colour at $M^{*}_{R}$ is $(B-R)_{*} = 1.485 \pm 0.009$ magnitudes, with a CMR slope of
$-0.031 \pm 0.006$.  For the fainter subsample of A0085, A0119 and 
A0496 combined, which extends to $R < -18.0$, the slope steepens
slightly to $-0.042 \pm 0.007$ with the normalized colour being $1.463
\pm 0.012$.  Note that \cite{LopBarYee04} find a CMR slope of
$-0.052\pm0.008$ for $M_R \le -18.6$, which is consistent with our result within the
uncertainties.

The ``1-sigma'' scatter in the colour-relation is $0.07$ magnitudes, as
estimated via the SIQR.  Note that the measurement error in \br\ is
estimated to be $0.03$, so this yields a intrinsic scatter of $0.06$.  
This is consistent with the value ($0.074$) found by
\cite{LopBarYee04}, although their limiting magnitude ($M_R = -18.6$)
is somewhat fainter whereas their radial limit ($0.2\, r_{200}$) is
smaller than ours ($0.5 \, r_{200}$).

\subsection{Fraction of Blue Galaxies}
The blue fraction was defined by \citet[hereafter BO]{ButOem84} to be
the fraction of galaxies brighter than $M_V = -20$, lying
within a third of the cluster radius and bluer than the red sequence
by 0.2 mags in $B-V$.  
After correcting for cosmology, we find that their magnitude corresponds to 
$M_R = -19.1$, which is only $0.4$ mags fainter than our completeness limit. 
To match the BO definition as closely as possible, here we define a blue galaxy is as one with $(B-R) < 1.25$ after having normalized to $M_{*}$ by correcting for the CMR slope, i.e. 0.24 magnitudes bluer than the red sequence. Within $0.5\, r_{200}$ there are only 35 blue galaxies of 595 in total in the 8 clusters, yielding a blue fraction of $5.9\pm1.0$\%.  For A0085, A0119 and A046, the clusters complete to $M_R < -18$, within $0.5\,r_{200}$, the blue fraction climbs from $12/183$ ($6.5\pm1.9\%$) at $M_R<-19.5$ to $27/259 = 10.4\pm2.0$\% at $M_R < -19$. This suggests that the blue fraction for the whole sample to $M_R < -19$ is $10\pm2\%$.  This blue fraction is considerably less than the blue fractions ($\sim$25\%) at the same radius quoted in \cite{AndQuiTaj06} for clusters at $z \sim 0.35$.

\subsection{Colour-Magnitude Relations for Individual Clusters}

\begin{figure*}
  \includegraphics[width=\textwidth]{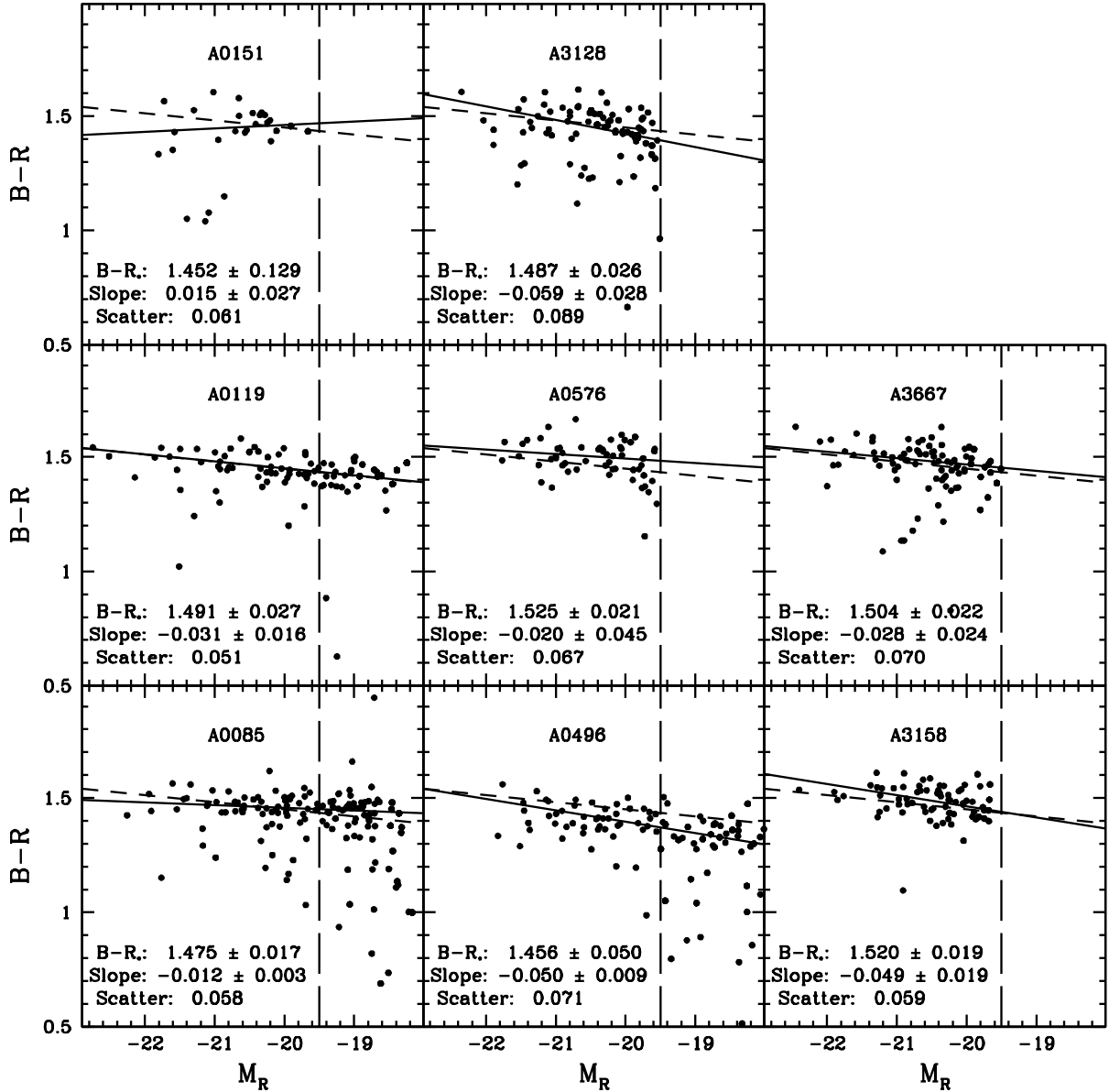}
\caption{ The colour-magnitude diagrams for each of the clusters. The
  solid line indicates the least-squares fit for the data in the
  cluster, and the dashed line shows global colour-magnitude relation.
  The fits are to the $M_R < -19.5$ sample.
\label{fig:cluscolmag}}
\end{figure*}

The individual cluster colour-magnitude relations are shown in Figure
\ref{fig:cluscolmag}.  The normalized colours $(B-R)_{*}$ for each cluster are consistent with the global value. 
Only the slope in A0085 deviates by more than $2\sigma$ from the global slope.  
Thus there is no evidence for deviation from the universal CMR in our sample.

\subsection{Bulge and Disc Colour-Magnitude Relations}

Having established that the total colours are uniform and consistent
with previous results, we now turn to the main goal of this paper.

\begin{figure*}
\includegraphics[width=\textwidth]{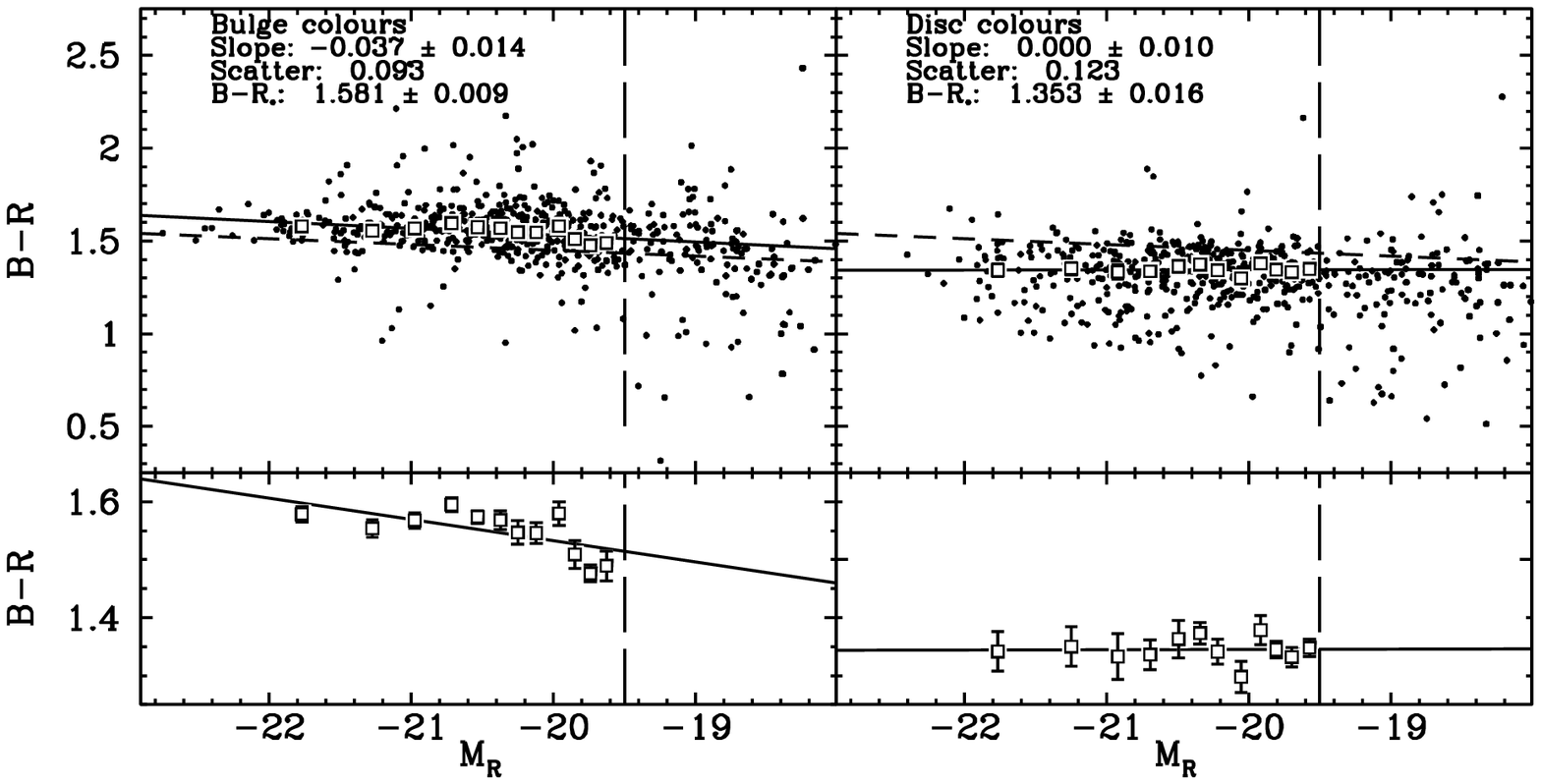}
\caption{ (Left Panel) Bulge colour vs.\ total $R$-band magnitude.
  All galaxies which are disc-dominated $(\bt < 0.2)$ have been
  removed.  Symbols are as in \figref{cluscolmag}. The solid line is
  the robust fit to bulge colours with $\rp < r_{200}/2$ and $M_R <
  -19.5$. The dashed line is the total-colour-magnitude relation for
  comparison. (Right) Disc colour vs.\ total $R$-band magnitude.  All
  galaxies which are bulge dominated $(\bt > 0.8)$ have been removed
  for this comparison. Symbols and curves are as in the left panel.
\label{fig:bulgedisccolmag}
}
\end{figure*}

Our two-band GIM2D fits yield the colours and CMRs of the bulge and
disc components separately. However, galaxies with very low \bt\ do
not have reliable bulge colours, so, in measuring the bulge
colour-magnitude relation, we have omitted galaxies with $\bt < 0.2$.
This culled bulge colour-magnitude diagram is shown in the left panel
of Figure \ref{fig:bulgedisccolmag}.  Note that the horizontal axis is
the total $R-$band magnitude.  The slope of the bulge CMR, $-0.037 \pm
0.014$, is very similar to the slope of total-light CMR, however the
normalized colour, $1.581 \pm 0.091$, is 0.1 magnitudes redder than
the total colour.  Using the population synthesis models of Maraston,
for an assumed metallicity $[Z/Z_{\sun}] = 0.3$ dex, this colour
corresponds to an SSP age of 12 Gyr. At fainter mags $M_R < -18$ the
bulge CMR steepens to $-0.067\pm0.017$.

The measured scatter in the bulge CMR ($0.093$) is actually somewhat
larger than the total CMR scatter ($0.07$).  This can be understood if
one allows for additional error in the bulge-disc decomposition
itself: some disc light is incorrectly assigned to the bulge and vice
versa. In \secref{data}, we estimated from repeat measurements that the measurement error on the bulge colour is $\sim 0.07$, and so the intrinsic scatter is likely to be much lower.

If we fit bulge colour as a function of bulge magnitude, we obtain a flatter slope $-0.025 \pm  0.013$, and larger scatter (0.098).However, in \secref{discusscmr} we will show that bulge colour is most closely linked not to total magnitude but rather to central velocity dispersion.

The disc colour-magnitude relation for galaxies with $\bt < 0.8$ is
shown in the left panel of Figure \ref{fig:bulgedisccolmag}.  We find
that discs have a flat CMR: the measured slope is $0.000 \pm 0.010$.
The disc colour for an $L_*$ galaxy ($1.353 \pm 0.016$) is
significantly bluer than that of the bulge component at the same total
magnitude.  Thus, in clusters, discs are bluer than bulges by $0.228
\pm 0.024$ mags at $L_*$.  The bulge-disc colour difference is larger than, but marginally
consistent with, that found by \cite{PelBal96} for S0-Sbc galaxies in the field ($0.045\pm0.097$).  On the other hand, the colour difference in clusters is smaller than the value  $0.301\pm0.014$ for a S0/a-Sdm field sample \citep{MacCouBel04}.  
At fainter magnitudes, the difference in colour is much less: at $M_R = -18$, $\sim 0.09\pm0.06$ mags.  The colour-magnitude relations are summarized in \tabref{colmag}. When disc colours are regressed against disc magnitude the slope is slightly negative, although the difference from zero is not statistically significant: $-0.006 \pm  0.015$, and the scatter is similar ($0.120$) to the scatter versus total magnitude..

\begin{figure}
\includegraphics[width=\columnwidth]{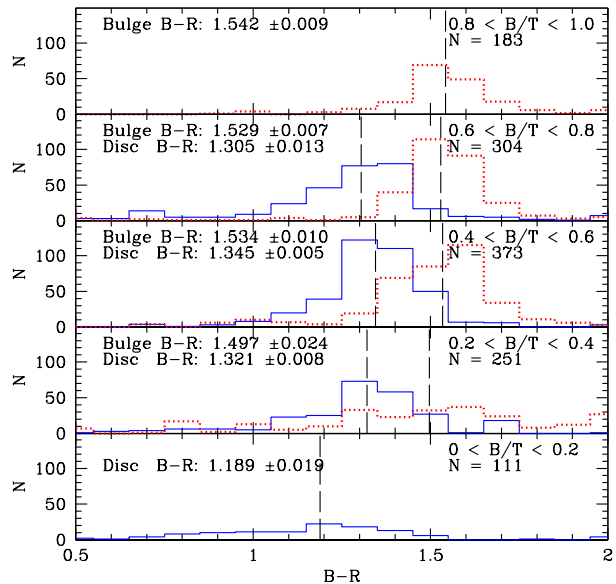}
\caption{Histograms of the colours of disc (solid lines, for galaxies
  with $\bt < 0.8$) and bulge (dotted lines, for galaxies with $\bt >
  0.2$) colours, for subsamples with different \bt.  Note there is little dependence of disc or
  bulge colour on $\bt$, except for discs in very late types ($\bt <
  0.2$).
  \label{fig:discbulgehist} }
\end{figure}

It is important to investigate whether bulge and disc colours depend
on morphology. For example, in the field, \cite{MacCouBel04} 
found that the colours of both bulges and discs are correlated with morphology
 in the sense that later morphological types host both blue bulges and blue discs. 
Figure \ref{fig:discbulgehist} shows a histogram of the
bulge and disc colours for different bins of bulge-to-total ratio. The
bulge and disc colours show no significant dependence on morphology for
$\bt > 0.2$.  Discs in $\bt < 0.2$ galaxies are significantly bluer
than the discs in higher \bt\ galaxies. However, there are few ($9\%$)
such low $\bt$ galaxies in our sample.

\begin{table}
\caption{Galaxy colours and properties as a function of magnitude.}
\label{tab:mag}
\begin{tabular}{@{}lcr@{$\pm$}lr@{$\pm$}l}
\hline
Property &
Mag Limit. &
\multicolumn{2}{c}{slope} & 
\multicolumn{2}{c}{value at $L_{*}$} \\
 \hline
Total Colour & $M< -19.5$ & $-0.031 $ & $  0.006$ & $ 1.485 $ & $ 0.009 $ \\
Bulge Colour & $M< -19.5$ & $-0.037 $ & $  0.014$& $ 1.581 $ & $  0.009 $ \\
Disc Colour & $M< -19.5$& $ 0.000 $ & $  0.010$& $ 1.353 $ & $  0.016 $ \\
$B/T$ &   $M< -19.5$ &  $ -0.069 $ & $  0.020 $ & $  0.614 $ & $  0.020$ \\
\\
Total Colour & $M< -18$ & $-0.042 $ & $  0.007 $& $ 1.463 $ & $  0.010 $ \\
Bulge Colour & $M< -18$& $-0.067 $ & $  0.017 $& $ 1.596 $ & $  0.016 $\\
Disc Colour & $M< -18$& $ 0.000 $ & $  0.014$& $ 1.307 $ & $  0.029 $ \\
$B/T$ & $M< -18$ & $ -0.069$ &  $0.019 $ & $ 0.643$ & $0.026$ \\
\hline
\label{tab:colmag}
\end{tabular}
\end{table}

\subsection{Deconstructing the Colour-Magnitude Relation: Contributions
  from Bulge and Disc colours  and from the Bulge-to-Total ratio}

From the bulge and disc colour magnitude diagrams, it appears that the
global CMR slope is due primarily to the colour of the bulge
component.  A quantitative check of this conclusion can be performed
by reconstructing the global slope using the bulge and disc CMRs and
the dependence of \bt-ratio on magnitude.  Specifically, the global
colour-magnitude relation can be decomposed as
\begin{equation}
  \begin{split}
    \frac{\del (B-R)}{\del M_R} 
    & = \frac{\partial \bt}{\partial M_R}[\expec{B-R}_B - \expec{B-R}_D] 
\notag \\
    & + \frac{\partial (B-R)_B}{\partial M_R}\expec{\bt} \notag \\
    & + \frac{\partial (B-R)_D}{\partial
      M_R}(1-\expec{\bt})
\end{split}
\end{equation}
For our $M_R < -19.5$ sample we measure an average $\expec{B/T} =
0.566\pm0.04$ and an average difference in bulge and disc colour is
$\expec{B-R}_B - \expec{B-R}_D = 0.237\pm0.013$. 
Inserting the measured slopes from \tabref{colmag} and evaluating each term yields 
$-0.014\pm0.004$ for the first ($\bt$) term, 
$-0.021\pm0.008$ for the second (bulge CMR) term and zero from the third
(disc CMR) term.

Thus the origin of the ``tilt'' or slope of the total-light-CMR is due mostly (60\%) to the
slope of bulge CMR, with the remainder of the total-light-CMR slope arising from the change
in \bt\ as a function of magnitude, which mixes in increasing more bluer disc light at fainter magnitudes. 
These trends, however, do not give insight into the ages or metallicities of the stellar populations of our sample. We
address this topic in \secref{discusscmr} below.

\section{Discussion: The Origin of the CMR}
\label{sec:discusscmr}

We have shown that the tilt of the CMR is due primarily to the bulge CMR, and to a lesser extent, to the change in morphology along the red-sequence.  The bulge-CMR itself has been thought to be due to an age-luminosity relation, a metallicity-luminosity relation, or both. While optical colours alone cannot break the ``age-metallicity'' degeneracy, in principle it is possible to do so by studying spectroscopic absorption linestrengths.  \citet{SmiLucHud09c} and \citet{GraFabSch09} have shown that the dominant ``driver'' of stellar populations is the central velocity dispersion, $\sigma$, and that any dependence on, for example, stellar mass is secondary.

\begin{figure*}
  \includegraphics[width=\textwidth]{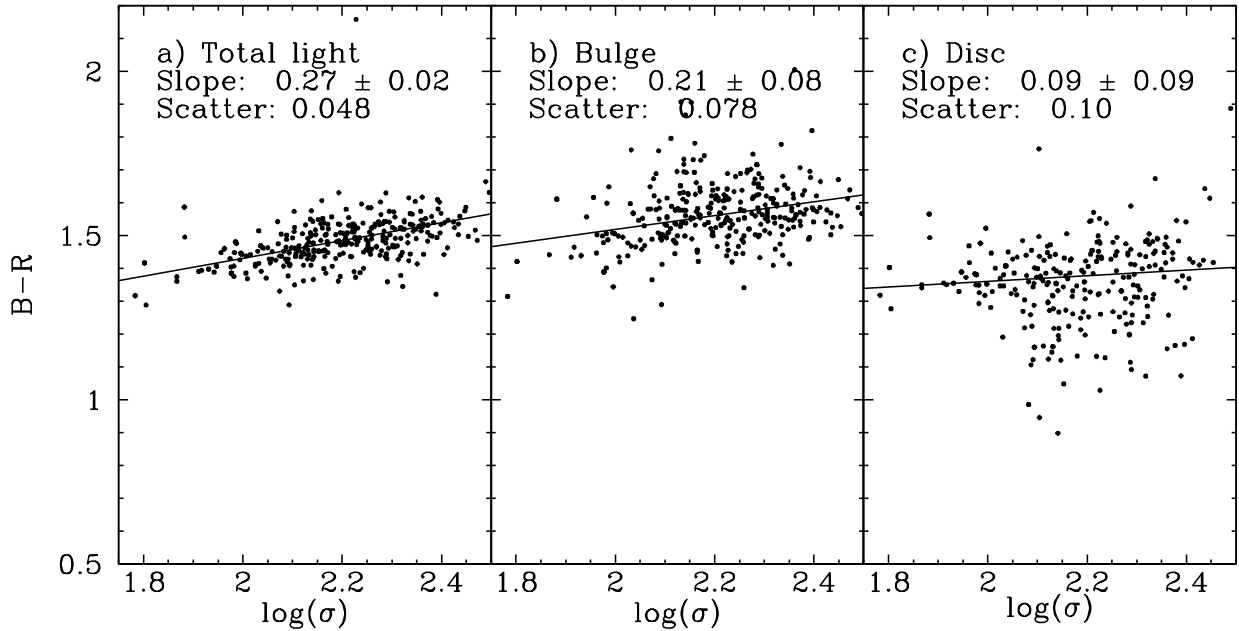}
  \caption{ Colour versus $\log(\sigma)$ for galaxies in the NFPS-II sample (red and emission free), with otherwise the same selection criteria as in Figures \ref{fig:colmag} and \ref{fig:bulgedisccolmag}. The three panels show the total
    colour, bulge colour and disc colour, from left to right. 
    \label{fig:colsig}}
\end{figure*}

It is therefore interesting to examine the colours as a function of velocity dispersion, and to model these colours as a function of stellar population parameters.  Using the NFPS velocity dispersions \citep{SmiHudNel04} for a subsample of emission-free, red galaxies with $M_R < -19.5$, we obtain the colour-$\sigma$ relation (CSR) shown in \figref{colsig}; the slope is $0.27 \pm 0.04$, consistent with previous determinations \citep{MatGuz05}. The bulge-CSR has a statistically-significant slope of $0.21\pm0.08$, whereas that of the disc is statistically insignificant ($0.09\pm0.09$).  We note in passing that the scatter in colour, for bulges and for discs, at fixed $\sigma$ is lower than the scatter at fixed magnitude (for the \emph{same} red-selected subsample), suggesting that velocity
dispersion and not, for example, stellar mass determines the colour of the disc as well as the bulge.

The flatness of the disc-CMR and the disc-CSR suggests that there is no significant variation in the stellar populations of the disc component along the sequence. As shown in the Appendix, the disc colour is consistent with an SSP age of 4.5 Gyr and solar metallicity.

It is possible to combine the disc colours, the GIM2D fits and nuclear linestrength data to derive ages and metallicities of the bulge component. These fits are described in detail in the Appendix, we summarize the results here:  the fits require a variation in bulge age, in the sense that bulges in lower velocity dispersion galaxies are younger, and a weaker variation in bulge metallicity.

An age variation along the bulge sequence is consistent with the results of \cite{SmiHudLuc06} from the NFPS. They included $B/T$ as an additional parameter in their fits of age, metallicity and \ale, and found that bulges are somewhat older than discs, by, on average, a factor 1.5, but that the effect of $B/T$ was weaker than the dependence on velocity dispersion. Thus, if discs are 4.5 Gyr old, one would expect bulges to have an age of $\sim 7$ Gyr. This is reasonably consistent with the fits described in the Appendix for which bulge ages ranging from 5 Gyr to 12 Gyr, as a function of velocity dispersion. The data are certainly consistent with scenario in which the bulge CSR (and CMR) ``tilt'' is largely due to age, with metallicity being a secondary effect.

\section{Environmental Trends within Clusters}
\label{sec:env}
        
It has long been known that galaxy properties are correlated with
environment, the most well-established trend being the
morphology-density relation \citep{Dre80}. While the dependence of
total colours on environment has been established by a number of
authors \citep{BalBalNic04, LopBarYee04, HogBlaBri04}, there has been
little attention paid to the colours of bulges and discs seperately.
In this section, we consider two indicators of environment: the
projected cluster-centric radius, and the local density. The former is
connected to the properties of the cluster as a whole, whereas the
latter is more sensitive to local structure.

\subsection{Dependence on Cluster-centric Radius}

\begin{figure}
\includegraphics[width=\columnwidth]{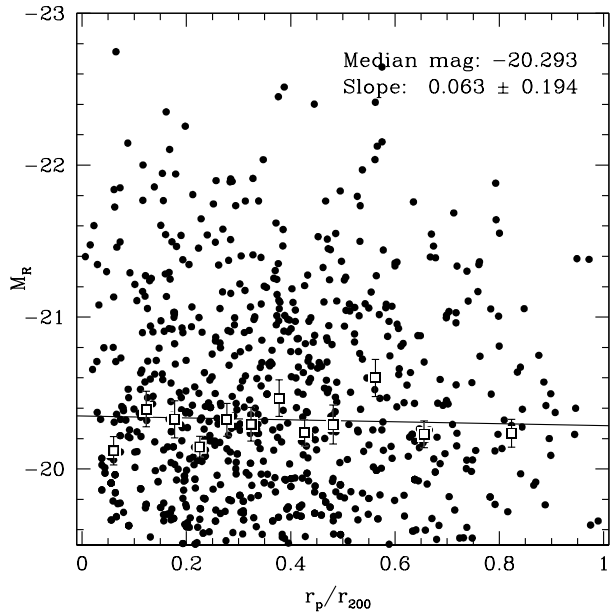}
\caption{ 
$R-$band absolute magnitude as a function of projected cluster-centric radius.
\label{fig:magrad}
}
\end{figure}

\begin{figure*}
  \includegraphics[width=\textwidth]{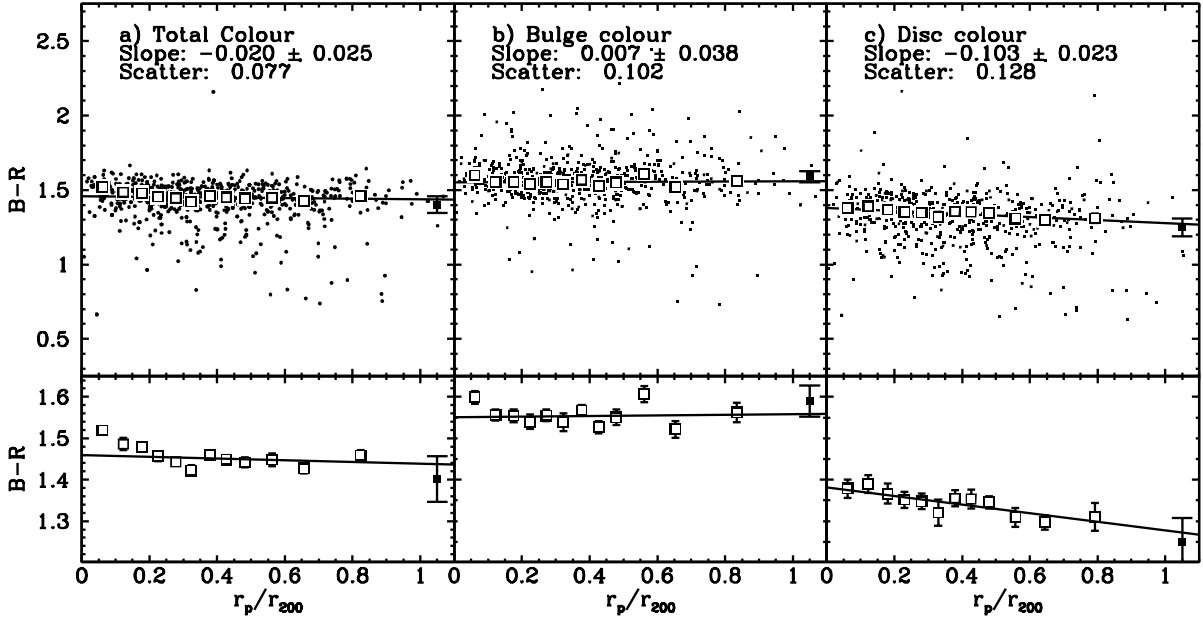}
\caption{ Total colour, bulge colour, and disc colour versus projected radius in units of $r_{200}$
  for the $M_R < -19.5$ sample.  The upper panels show individual points and medians (open squares).  The filled square denotes the median for the field sample, and is placed arbitrarily at 1.05 $r/r_{200}$. 
\label{fig:colrad}
}
\end{figure*}

Before turning to the dependence of colours on measures of environment, we must first examine the possibility of luminosity segregation, because, if there were significant luminosity segregation in clusters, one might expect to see a colour-radius effect arising from the combination of luminosity segregation and the CMR. \figref{magrad} shows magnitude as a function of radius.  The brightest galaxies, with $M_{R}<-22$ appear to prefer $r\sbr{p}/r_{200}<0.6$. For the remainder of the galaxies, no evidence for luminosity segregation in our sample. This result is in agreement with the recent studies  \citep{AdaBivMaz98,BivKatTho02,PraDrideP05,MerMerHai06,vonWilKau10}.  Furthermore, when if we subtract the CMR from the observed colours and look for the dependence of the colour residuals on radius, we find trends of similar significance to those found below. Hence for simplicity, we simply fit colour as a function of cluster-centric radius.

Figure \ref{fig:colrad} shows total, bulge and disc colours versus
projected cluster-centric radius in units of the virial radius:
$\rp/r_{200}$, Here, only the data with $M_R < -19.5$ are used.  The
left panel shows the total colour as a function of radius.  The slope
of the fit to median colours is $-0.020 \pm 0.025$.  The bulge colours
are shown in the middle panel.  The slope of the bulge-colour-radius
relation is not statistically significant.  Finally, the disc colours
are shown in the right-hand panel, where the radial slope is
statistically significant at the 4.5$\sigma$ level: $-0.103 \pm
0.023$. These fits, as well as those described below, are summarized
in \tabref{radial}.

\begin{figure}
  \includegraphics[width=\columnwidth]{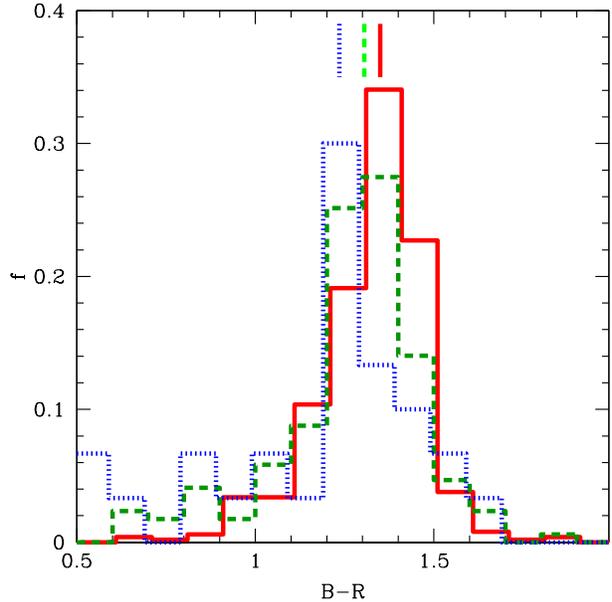}
\caption{ Histograms of disc colour for two radial ranges $\rp < 0.3  r_{200}$ (red solid) for $\rp > 0.3 r_{200}$ (green dashed) and in the field (blue dotted).  The vertical marks indicate the median colours of each sample.
\label{fig:disccolhist}
}
\end{figure}

Visual examination of \figref{colrad} may hint that much of the change
in disc colours is due to a tail of blue discs in the outskirts of
clusters.  However, recall that we are fitting the median colour (not
the mean) and this is much less sensitive to outliers.  The disc
colours for ``core'' ($\rp < 0.3 r_{200}$) and ``outskirt'' ($0.3 <
r_{p}/r_{200} <1$) samples are shown in \figref{disccolhist}. This
confirms that, while such a blue tail does exist, the peak of the
distribution is also shifted to the blue in the ``outskirts'' bin.
Moreover, this trend continues to the field: the peak of the colour
distribution shifts blueward, and the fraction of very blue discs
($\br < 1$) increases.  A Kolmogorov-Smirnov (K-S) test shows that the
``core'' bin is significantly different from both the ``outskirts''
bin and from the field, at 99.92\% and 99.5\% confidence levels
respectively.  \emph{Thus there is a clear difference between the
  cores of rich clusters and the field.}

We have also performed the same bulge and disc fits as a function of $\log(\rp/r_{200})$ (see \tabref{radial}). This functional form gives greater leverage to the innermost regions.  For this parametrization, the radial dependence of the total colour is highly significant: the slope is $-0.092 \pm 0.013$ per decade in radius.  Furthermore, with this parametrization, the bulge colour dependence is also statistically significant ($-0.046 \pm 0.018$), while the disc dependence ($-0.065 \pm 0.015$) remains significant. The difference between the results from the linear fit and the logarithmic fit arise from the ``core'' regions $\rp \lesssim 0.2\,r_{200}$, where bulge colours (and consequently the total colours) appear to be somewhat redder than at larger annuli.

\begin{table}
\caption{Galaxy colours and properties as a function of environment. Slopes that differ from zero at greater than $3\sigma$ significance level are shown in bold type.}
\begin{tabular}{llr@{$\pm$}l}
\hline
\multicolumn{1}{c}{$y$} &
\multicolumn{1}{c}{$x$} &
\multicolumn{2}{c}{slope} \\
\hline
$M_R$ & $r_p/r_{200}$ & $ 0.063 $ & $ 0.194$ \\ 
$B/T$ & $r_p/r_{200}$ & $ 0.004 $ & $  0.061$ \\ 
Total Colour & $r_p/r_{200}$ & $-0.020 $ & $  0.025 $\\
Bulge Colour & $r_p/r_{200}$ & $ 0.007 $ & $  0.038 $\\  
Disc Colour & $r_p/r_{200}$ & $\mathbf{-0.103} $ & $\mathbf{0.023}$\\
 \\
$M_R$ & $\log(r_p/r_{200})$ & $ -0.182 $ & $  0.113$ \\
$B/T$ & $\log(r_p/r_{200})$ & $\mathbf{-0.108}$ & $ \mathbf{0.033}$ \\
Total Colour & $\log(r_p/r_{200})$ & $\mathbf{-0.092}$ & $ \mathbf{0.013}$ \\       
Bulge Colour & $\log(r_p/r_{200})$ & $ -0.046 $ & $  0.018$ \\
Disc Colour & $\log(r_p/r_{200})$ & $\mathbf{-0.065}$ & $ \mathbf{0.015}$ \\                
\\
$M_R$ & $\log(\Sigma_5)$ & $ 0.054 $ & $  0.064$ \\
$B/T$  & $\log(\Sigma_5)$ & $0.048$ & $0.019$\\
Total Colour & $\log(\Sigma_5)$ & $\mathbf{0.032}$ & $ \mathbf{0.008}$\\
Bulge Colour & $\log(\Sigma_5)$ & $ 0.006 $ & $  0.014 $\\
Disc Colour & $\log(\Sigma_5)$ & $\mathbf{0.050}$ & $ \mathbf{0.013}$\\
\hline
\end{tabular}
\label{tab:radial}
\end{table}

We have also attempted to extend this analysis to fainter magnitudes but found
that, when the sample is restricted to only the three clusters
complete to $M_R < -18$, the error bars on the slopes are considerably
(a factor 2-3 times) larger and so no firm conclusions can be drawn.

\subsection{Density as an Environment Indicator}

The cluster-centric radius is a reasonable proxy on average for the depth of the cluster potential well, but in reality clusters are built from groups and singles \citep[see][for a detailed discussion.]{McGBalBow09}. If the environmental influences are attributable to a given galaxy's previous (``group-like'') environment (``pre-processing'') then an environmental indicator based on the local density might be more sensitive to such effects.

To test this we define the local density $\Sigma_5$ as follows.  Using galaxies brighter
than our absolute magnitude limit of $-19.5$, we find the distance to
the fifth nearest neighbor in the cluster and compute the density
using this radius.  It turns out that this definition is almost
identical (within a few hundredths of a magnitude) to the limit used
by \citet{BalBalNic04}, who used SDSS $r$ mags. However, rather than
defining the density within a $\pm1000 \kms$ redshift slice, we simply
consider only galaxies that are clusters members.

\begin{figure*}
  \includegraphics[width=\textwidth]{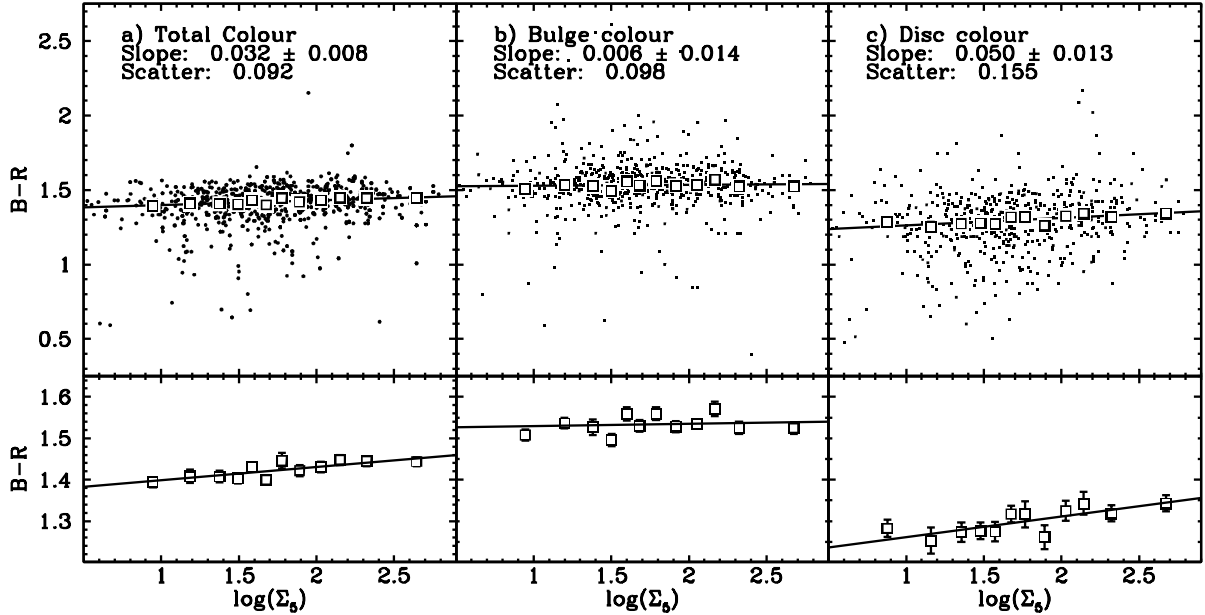}
  \caption{Total colour, bulge colour, and disc colour versus projected
    density. \label{fig:logden}}
\end{figure*}

As seen in Figure \ref{fig:logden}, the total colour varies as $0.032 \pm 0.008$ per dex in density, in the
expected sense that galaxies in higher density environments are
redder.  The bulge colour slope is not statistically significant
($0.006 \pm 0.019$) but the disc colour slope is $0.050 \pm 0.013$,
significant at the 3.9 $\sigma$ level.  
The significance of the trends with density are
similar to, or slightly lower in significance than the radial trends as a function of $\log(\rp/r_{200})$, suggesting that these two measures of environment are largely degenerate when applied to rich clusters.

\section{Discussion}
\label{sec:discuss}

\subsection{Bulges}

The bulge colours have show no dependence on environment from the cores of rich clusters to the field. At face value, this is compatible with the results from morphologically-selected samples. For  example, \cite{vanFraKel98} found no clustercentric variation in their E sample (although they did find a trend for their S0 sample). \cite{HogBlaBri04} found a very weak dependence of the peak of the colour distribution of S\'{e}rsic $n > 2$ galaxies with density. Finally, in a cluster at $z=0.83$, \cite{KooDatWil05} also found that bulge colours were insensitive to cluster-centric radius. 

However, we note that it is possible for the stellar populations of bulges to change in such a way as to preserve $B-R$ colour. For this to occur, changes in metallicity and age would have to be coordinated such that $(\Delta Z)/Z= -0.7 (\Delta \mathrm{Age})/\mathrm{Age}$ \citep{Wor94}. In fact,  an anti-correlation with exactly such a slope is indeed observed from spectroscopic analyses which break the age-metallicity degeneracy \citep{TraFabWor00, SmiLucHud08}. So it is possible that there is greater variation in the stellar populations of bulges than their colours suggest.

\subsection{Disc fading}

The dependence of disc colour on radius strongly suggests a scenario in which their star formation has been shut off by the cluster environment, after which the discs fade and redden. It is interesting to determine whether this process is abrupt, as one might expect if the \emph{cold gas} is stripped by ram pressure, or slower, as might be expected in the ``strangulation'' scenario \citep{LarTinCal80}, in which the hot gas halo is stripped, but the cold gas remains to fuel star formation. The former scenario certainly exists:  \cite{SmiLucHam10} found that at least 1/3 of blue galaxies in the 1 \hmpc-radius core of Coma show UV tails extending from the disk suggestive of star formation during ram pressure stripping, and the direction of the tails suggests that the stripping is occurring on infall. It is not clear, however, how efficient and how rapid this process is. Evidence for a slower ``strangulation'' mechanism is favoured by a comparison of models with red and blue fractions \citep{ FonBowMcC08, vanAquYan08, WeiKauvan09, McGBalBow09}.
 
We have found that the difference in median colour between discs at the centre of the cluster and those at the virial radius is $0.103\pm0.023$ magnitudes in \br.  It is interesting to see if this colour difference can be reproduced with simple infall models in which star formation is truncated as the disc falls into the cluster. We use the simulations of \cite{GaoWhiJen04} to associate a given cluster-centric radius with a median infall time.  Their Figure 15 suggests that while objects in the centre of the cluster were accreted at $z=1$ (a lookback time of 7.7 Gyr), those at the virial radius typically were accreted at $z \sim 0.4$ (4.3 Gyr ago). 

If we assume that solar metallicity discs began forming stars at 10 Gyr ago at a continuous rate until crossing the cluster virial radius, followed by abrupt quenching,  we obtain a colour of $1.43$ for the disc in the centre of the cluster, in reasonable agreement with the median observed value of $1.38$.  However, at the virial radius, the predicted colour is $1.40$, which is quite red compared to the observed colour of $1.28$. Put another way, the predicted colour \emph{difference} (a more robust quantity than the absolute colour) is $0.03$ whereas the observed value is $0.10\pm0.02$. If we instead adopt a ``strangulation'' scenario for the disc in which star formation is not cut off abruptly, but instead declines with an exponential decay rate of 1 Gyr after infall, the predicted colour difference between a disc at the centre and one at the virial radius is $0.07$.  This is in better agreement with the observed radial gradient in disc colour. 

However, the above naive interpretation is significantly complicated by projection effects. There are two factors at play: first, an intrinsic three-dimensional colour gradient will be \emph{flattened} when viewed in projection; second,  there will be contamination by infalling field galaxies that are outside the virial radius but which appear within the virial radius in projection (and which have sufficiently low velocities to pass the $\pm 3 \sigma\sbr{cl}$ clipping). These field interlopers are found in the outer regions and so will \emph{steepen} the observed colour gradient. The fraction of such interlopers is difficult to determine. \cite{MamBivMur10} estimate that  the fraction of such interlopers is a strong function of projected radius. More specifically, for the subsample with  $r_{p} >  0.3 r_{200}$, one might expect $\sim 33\%$ of objects to lie beyond the virial radius in three dimensions (G. Mamon, private communication).  As a further complication, most of these ``interlopers'' will be infalling blue galaxies, but some may be ``backsplash'' galaxies which have passed through the centre of the cluster and have exited the virial radius. 

We can test the null hypothesis that there is no intrinsic variation within the cluster virial radius. Specifically, we simulate the colour distribution of the outer radial bin with a mixture of the disc colours from the ``core'' bin (67\%) with 33\% contamination from the field discs. While the median colour of this simulated population is slightly redder than the observations, a K-S test shows that there is no significant difference between it and the observed distribution of disc colours in the outer radial bin.  Thus we cannot rule out the hypothesis that the reddening begins beyond the virial radius. Nevertheless, the data do not exclude some reddening within the cluster itself: indeed a better fit is obtained if we assume that, intrinsically, the core galaxies are redder than the outskirt galaxies by 0.03 mag (in rough agreement with the ``abrupt quenching'' model discussed above). In summary, given the uncertainties associated with the interloper contamination fractions, and with the small size of the comparison field sample, it is difficult to decontaminate the cluster samples with sufficient accuracy in order to determine which scenario (abrupt quenching or strangulation) is at play. Nevertheless, we emphasize that there is indeed a difference between the disc colours of cluster-core galaxies and those at larger radii.

\subsection{Implications for the Morphology-Density Relation}

In the previous section, we showed that the data are consistent with a scenario in which the disc component fades within the cluster environment. Is this disc fading consistent with the buildup of the morphology-density relation? More specifically, are S0s formed by spirals after their discs fade and their spiral density waves are damped? 

To address this question, we first need to estimate the amount of fading. Comparing a hypothetical field spiral where the disc continues forming stars at a constant rate to one which was truncated 6 Gyr ago, the models of \cite{Mar05} predict a difference of a factor of $\sim 2.5$ in the $R$-band stellar mass-to-light ratio and a factor $\sim 4$ in the $B$-band. Thus, assuming the bulge is unaffected by environment, as a galaxy's disc fades, the total $R$ luminosity will drop by a factor less than 2.5, and \bt\ will increase by a factor up to 2.5.  General support for this picture comes from studies \citep{NeiMaoRix99,BedAraMer06} which have shown that the S0 \citet[hereafter TF]{TulFis77} relation is approximately consistent with being a faded version of the spiral TF relation. 

However, there are problems with applying this simple fading scenario to map the field population into the cluster population as a whole. One problem arises when comparing the cluster \bt\ distribution to that in the field. In \figref{bthist}, we showed the field $B$-band \bt\ distribution from \cite{AllDriGra06}, which has a large excess of late types with $\bt < 0.1$ (and many with no bulge at all).  As noted above, in the $B$-band, the fading might be as much as a factor of 4 between the field and the cluster. Thus field late-type spirals which would have had $\bt < 0.1$ had their star formation continued, might appear as $\bt \sim 0.2$ if their star formation was truncated at cluster infall. However, \figref{bthist} does \emph{not} show an excess of objects with $\bt \sim 0.2$, but rather a broad distribution peaking at $\bt \sim 0.5$ in the $B$-band.  Thus in order to create the cluster \bt\ histogram from that in the field, one needs to either transform the discs of late-type spirals into bulges, perhaps by ``harassment'' \citep{MooLakKat98}, or to remove the very low \bt\ objects from the distribution entirely, either by fading them below the magnitude limit or by destroying these galaxies.

A further problem with the scenario in which all S0s are formed from present-day spirals is the observation that the bulges of S0s are larger and brighter than those of spirals \citep{Dre80, BurHoHuc05, ChrZab04}. Therefore it seems that while the brighter S0s in present day clusters are unlikely to be the faded remnants of today's disc galaxies, this does not exclude fading as possible mechanism for creating \emph{lower-luminosity} S0s. Whether the brighter cluster S0s are consistent with being the faded versions of \emph{higher-redshift} spirals remains to be seen.

\section{Conclusions}

We have shown that the ``tilt'' of the \emph{total-light} CMR is driven
primarily by the tilt of CMR of the bulges (but not by the discs), and
to a lesser extent, by a change in the \bt\ ratio as a function of
magnitude along the red-sequence. The tilt of the CMR is therefore due
primarily to stellar population changes in the bulge as a function of
magnitude or $\sigma$.

In clusters, we find that discs in $L_*$ galaxies are $\sim 0.25$
magnitudes bluer than bulges. If we assume that discs have solar
metallicity, their colours are consistent with an SSP age of 4.5 Gyr.

Although bulge colours are insensitive to environment, the disc
colours show a statistically significant dependence on cluster-centric
radius in the sense that discs are bluer at the virial radius by $\sim
0.103\pm0.023$ magnitudes in \br\ compared to their counterparts of
the same total magnitude in the cluster centre.  This effect drives
the total-colour--radius relation noted by previous authors.  A
straightforward interpretation of this result is that it is due to the
quenching of star formation in the disc upon cluster infall.

In summary, the physical processes that affect star formation in
bulges are mass-dependent, and presumably primarily internal, whereas
star formation in the disc shows little dependence on mass, but a
strong dependence on environment as parametrized by cluster-centric
radius.

\section*{Acknowledgments}

We thank Gary Mamon for interesting discussions, and for calculating
contamination fractions for ``custom'' radial bins.  We are grateful
to Steve Allanson for help with his star formation history code, to
Lauren MacArthur for providing data in electronic form and to the
anonymous referee for useful comments and suggestions.

We gratefully acknowledge the substantial assignment of NOAO observing
resources to the NFPS program.  M. J. H. acknowledges support from the
NSERC of Canada. IRAF is distributed by the National Optical Astronomy
Observatory, which is operated by the Association of Universities for
Research in Astronomy, Inc., under contract with the National Science
Foundation.

\bibliographystyle{mn2e} 
\bibliography{mjh}

\appendix

\section{Ages and Metallicities of Bulges and Discs from Colours and Spectra}

Bulges and discs have different colours, presumably due to differing stellar populations. Due to degeneracies, it is difficult to measure both ages and metallicities from optical colours alone. Spectra allow one to break this degeneracy, and ideally, one would use IFU spectra to obtain spatially resolved ages and metallicties, but, at present, such IFU samples remain small and somewhat heterogeneous \citep[e.g.][]{RawSmiLuc08b, KunEmsBac10}. However, the NFPS contains spectra and stellar absorption linestrengths for 4097 red galaxies obtained from 2 arcsecond diameter fibres. From the linestrengths, a strong ``downsizing'' trend -- younger ages in less massive galaxies -- was found \citep{NelSmiHud05}. However, if the disc is younger than the bulge, and if \bt\ varies along the red-sequence, some of this trend may be due to variations in the fraction of central spectra arising from discs.

It is possible to use these spectra to infer bulge ages if we make some assumptions about the disc properties. Since the disc colour is independent of magnitude and $\sigma$, we will assume that discs have an SSP age of 4.5 Gyr, solar metallicty and solar ratios of $\alpha$-elements to iron. These parameters are chosen because they yield \cite{Mar98} model colours (assuming a \citep{Kro01} initial mass function), for the disc consistent with their median observed colour of 1.353. We will then model the contributions to the spectra from the bulge and disc components, and fit the stellar population parameters of the bulge using the central spectra. We will do this for subsamples binned by velocity dispersion, rather than magnitude because it has been shown that velocity dispersion, $\sigma$, is the driving parameter of the stellar populations \citep{SmiLucHud09c,GraFabSch09}.

\begin{figure*}
\begin{center}
\includegraphics[width=\columnwidth]{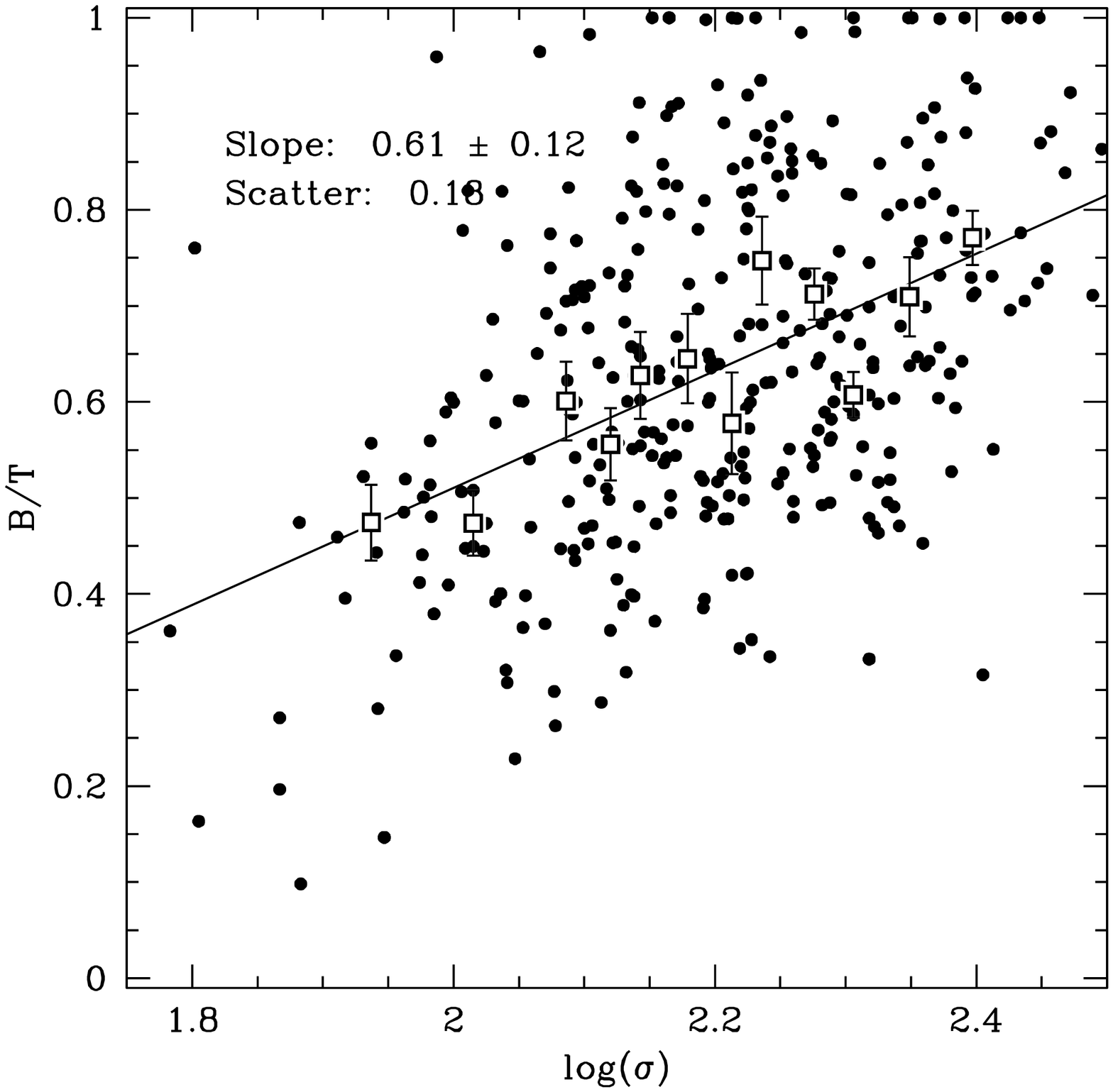}%
\includegraphics[width=\columnwidth]{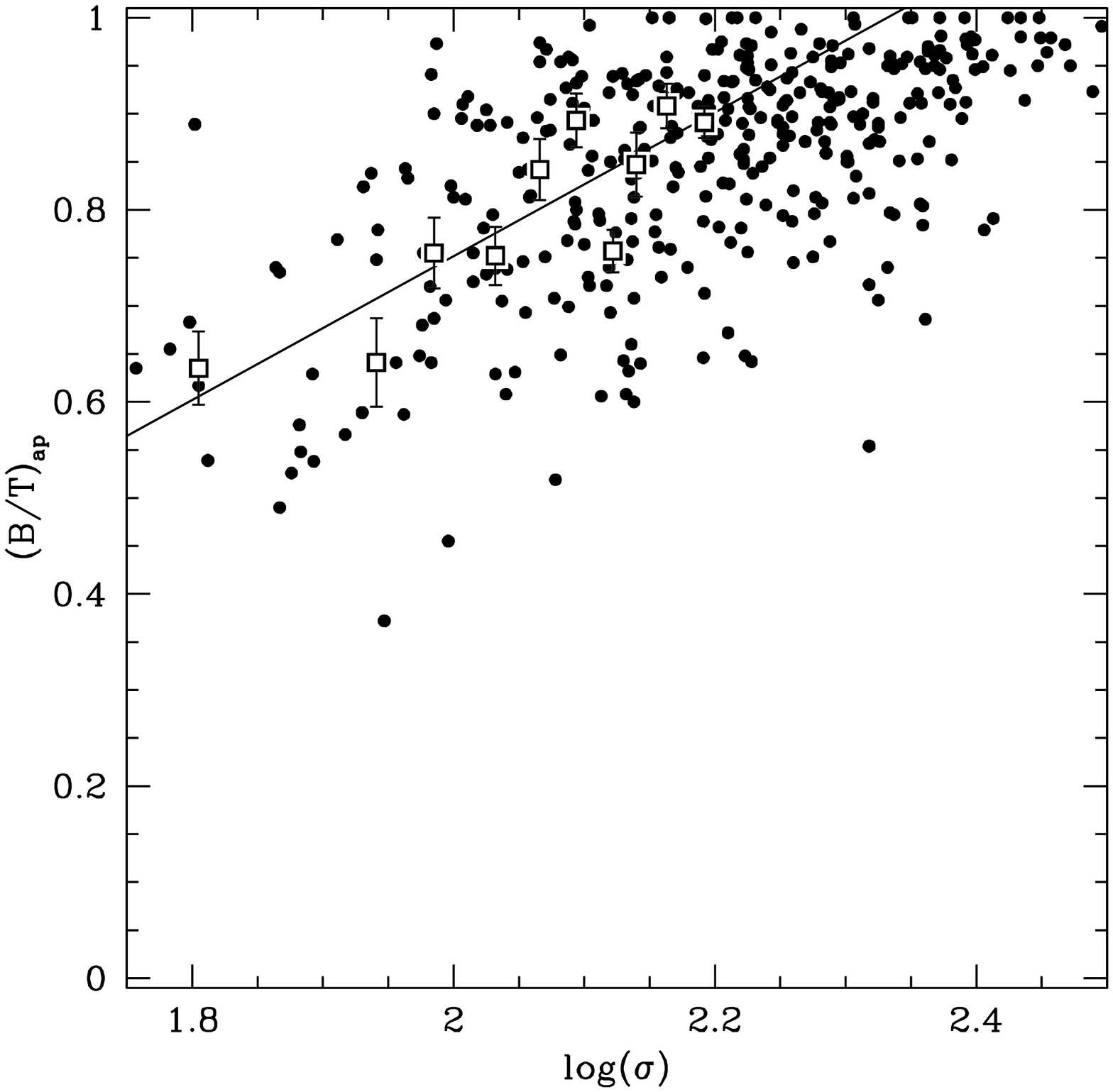}
\caption{$R-$band \bt\ as a function of velocity dispersion, $\sigma$. The left panel shows \bt, whereas the right panel shows the \bt$\sbr{ap}$, the \bt obtained from the light within a 2 arcsecond diameter aperture, chosen to match fibre spectroscopy. As in other plots, the solid squares show medians fo bins of $\sigma$. Note the tight correlation between $\log \sigma$ and \bt.}
\label{fig:btsig}
\end{center}
\end{figure*}

The central 2 arcsecond aperture contains a mixture of bulge and disc light, and so careful modelling is needed. In order to assess the contribution from bulge and disc to the central spectra, we use the GIM2D parameters to first calculate the bulge and disc light within a 2 arcsecond diameter aperture (matching the NFPS spectra), assuming that the exponential disc light can be extrapolated to the center.  We have found that, not only is $\sigma$ strongly correlated with stellar populations, it is also strongly correlated with morphology: \figref{btsig} shows \bt\ and the aperture \bt$_{ap}$, as a function of $\sigma$. Note the strong correlation and relatively low scatter (compare with the weak trend and high scatter in \bt\ versus magnitude in \figref{btrmag}).  Nevertheless, at all $\log (\sigma) > 1.8$, the central light is always dominated by the bulge. Thus for each $\sigma$ bin, we know \bt$\sbr{ap}$ from \figref{btsig}.

We proceed to model the ages and metallicities as follows. Given the disc age and metallicities, we vary the bulge age, metallicity and $\alpha$-enhancement. For each choice of bulge parameters, we calculate the stellar mass-to-light ratios of the bulge and disc components and, given the fraction of light in the bulge from $\bt_{ap}$, we solve for the stellar mass fraction in the bulge and in the disc components.  We use the composite stellar population models of \cite{AllHudSmi09}, which in turn are based on \cite{ThoMarBen03,ThoMarKor04},  to calculate spectral linestrength indices for the two-component model and compare these to the mean NFPS linestrengths for the same $\sigma$ bin, and then minimize to find the best-fit bulge age and metallicity. 

\begin{figure*}
\begin{center}
\includegraphics[width=\columnwidth]{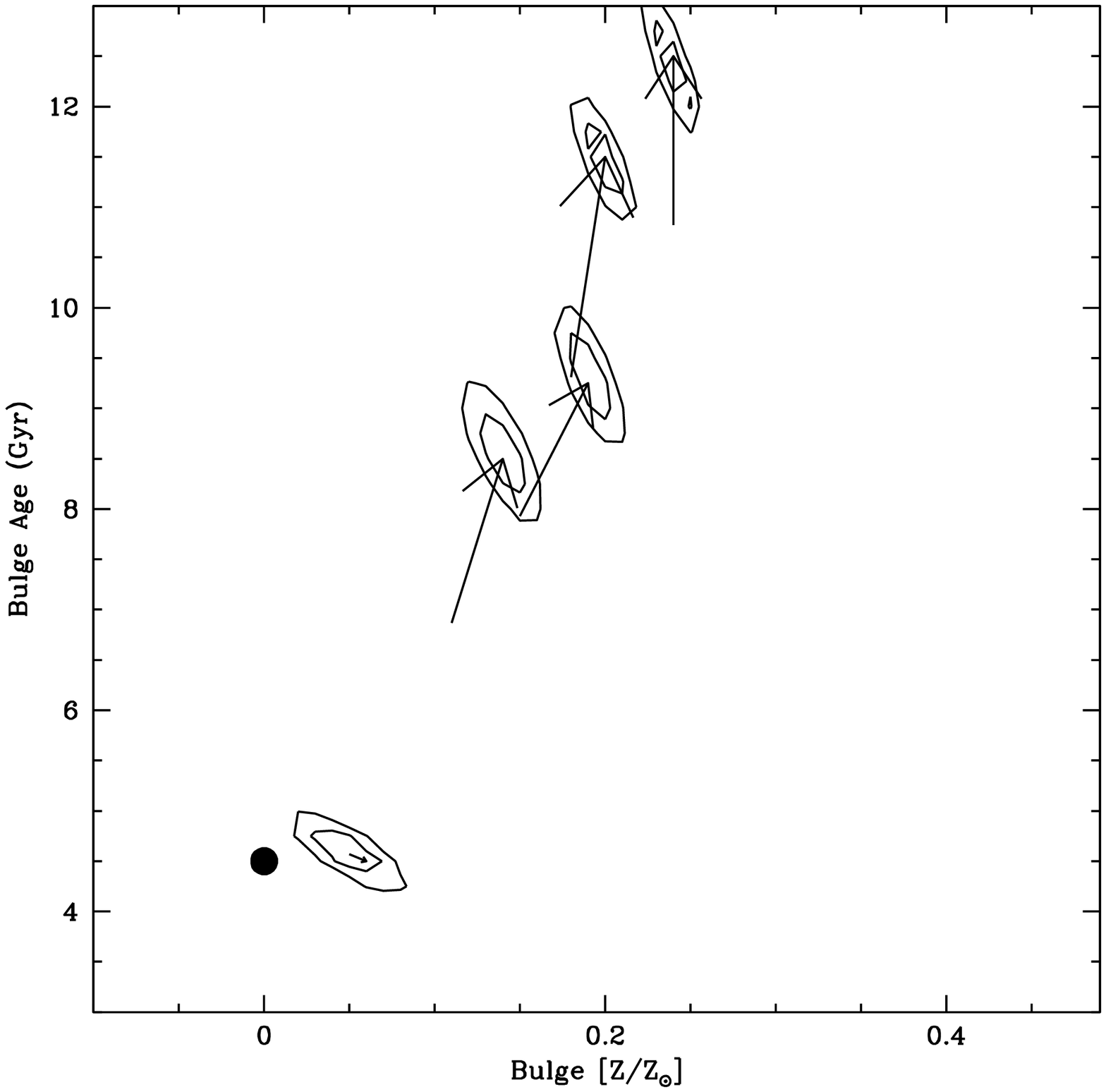}%
\includegraphics[width=\columnwidth]{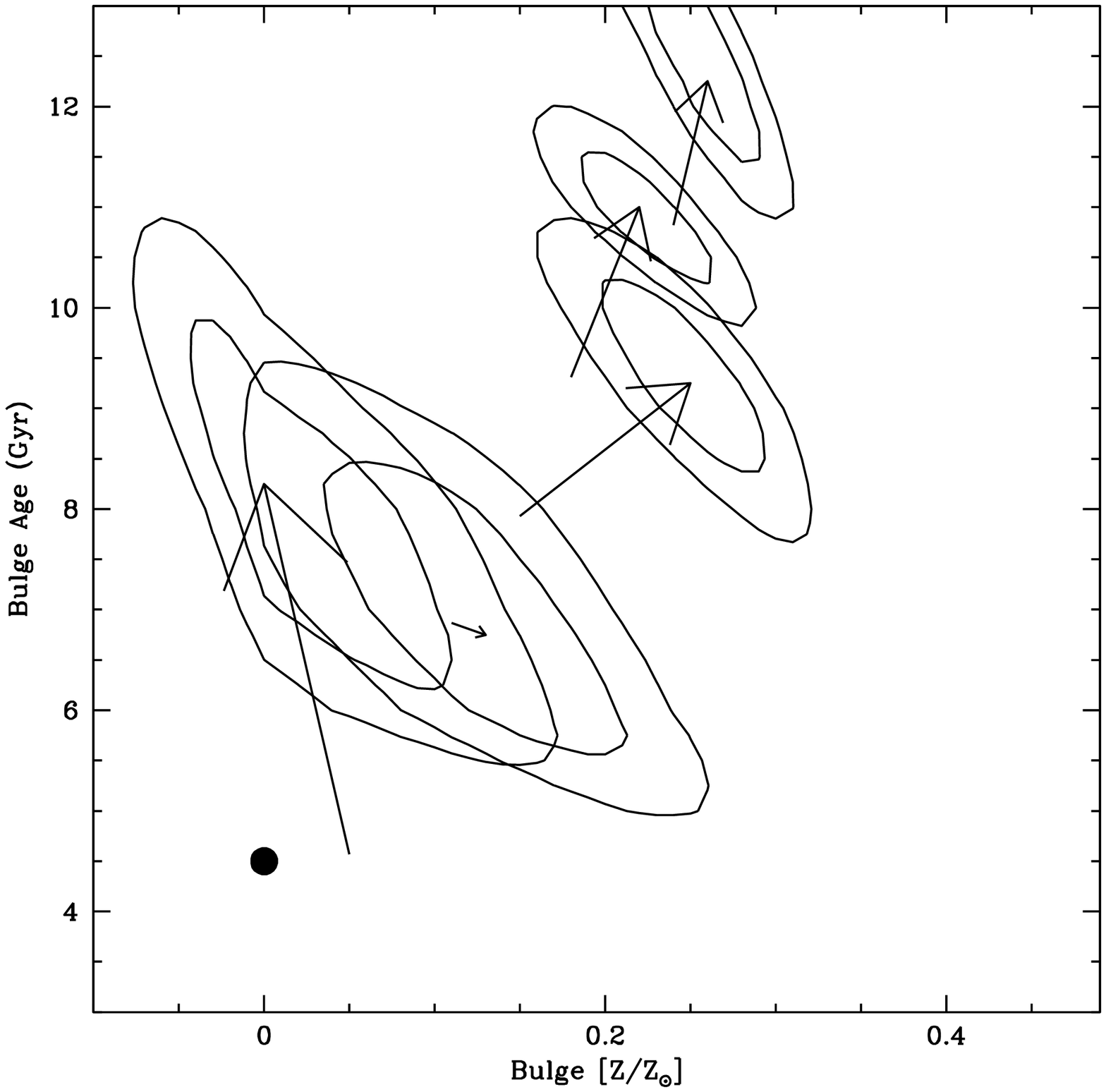}
\caption{%
Bulge age as a function of bulge metallicity. The left panel shows fits to the NFPS central linestrength data, assuming a 4.5 Gyr solar metallicity disc (indicated by the solid circle). Contours indicate 1 and 2 $\sigma$ confidence levels on two parameters jointly for bins with $\langle\log \sigma\rangle = 1.94,2.09,2.17,2.26,2.39$ from bottom left to top right. The base of each arrow shows the SSP parameters if we assume only a single component SSP as in \citet{NelSmiHud05}, 
whereas the tip of the arrow shows the best fit bulge parameters of the bulge+disc model.  The right panel is the same except for the data from 3 clusters in the Shapley Concentration from 
\citet{SmiLucHud07}, 
in comparison to single-component SSPs from 
\citet{AllHudSmi09} 
for bins $\langle \log \sigma \rangle =1.71,1.84,2.02,2.16,2.31$, from leftmost to upper right.  Note that the central spectra require that the bulge component of lower $\sigma$ galaxies is younger than that found in larger $\sigma$ galaxies (bulge ``downsizing'').}
\label{fig:bulgeagemetallicity}
\end{center}
\end{figure*}

The resulting bulge ages and metallicities as a function of $\sigma$ are shown in \figref{bulgeagemetallicity}.  We show results for two data samples: the NFPS \citep{NelSmiHud05}; and 3 clusters in the Shapley concentration \citep{SmiLucHud07}, which are also observed through a 2 arcsecond fibre, and are at a similar distance to the NFPS clusters studied in this paper.  In both cases, the effect of including a 4.5 Gyr disc component is to shift the bulges to older ages, typically by $1-2$ Gyr, compared to simple single-component SSP fits. There is some disagreement between the two samples for the lowest $\log \sigma \sim 1.8$ bin, for which the NFPS requires bulge ages $\sim 4.5$ Gyr, whereas the Shapley sample prefers ages $\sim 6.5$ Gyr. For both samples, the ``downsizing trend'', in which bulges with larger velocity dispersions are older, is present in bulge components.  

Of course, this result is based on the assumption that disc surface brightness profile continues to be an exponential right to the centre of the galaxy. If the surface brightness profile flattens towards the centre, as is common in S0's \citep{Fre70}, then the disc contribution is reduced compared to the bulge, and our conclusion regarding bulge downsizing is strengthened. It is, of course, in principle also possible that disc age does vary along the sequence, and that metallicity and/or dust conspire to achieve a disc colour which is independent of magnitude or $\sigma$,  but such scenarios would require fine-tuning and so seem rather contrived. Thus we conclude that the variation in bulge metallicity along the red-sequence appears to be relatively less important than bulge age in driving the bulge CSR (and hence CMR).

\end{document}